\documentclass[12pt,preprint]{aastex}
\usepackage{url}
\def\lea{\mathrel{<\kern-1.0em\lower0.9ex\hbox{$\sim$}}}
\def\gea{\mathrel{>\kern-1.0em\lower0.9ex\hbox{$\sim$}}}

\slugcomment{Submitted to The Astrophysical Journal}

\shorttitle{UVOT Stars: M67}
\shortauthors{Siegel et al.}

\begin{document}

\title{The {\it Swift\/} UVOT Stars Survey. II. RR Lyrae Stars in M~3 and M~15}

\author{Michael H. Siegel\altaffilmark{1}, Blair L. Porterfield\altaffilmark{1,2}, Benjamin G. Balzer\altaffilmark{1},
Lea M. Z. Hagen\altaffilmark{1,3}}

\altaffiltext{1}{Pennsylvania State University, Department of Astronomy, 525 Davey Laboratory,
University Park, PA, 16802 (siegel@astro.psu.edu, blp14@psu.edu, bgb5080@psu.edu, lea.zernow.hagen@gmail.com)}
\altaffiltext{2}{Current Address: Space Telescope Science Institute, 3700 San Martin Drive, Baltimore, MD, 21218}
\altaffiltext{3}{The Institute for Gravitation and the Cosmos, The Pennsylvania State University, University Park, PA, 16801}

\begin{abstract}
We present the first results of an near-ultraviolet (NUV) survey of RR Lyrae stars from the Ultraviolet Optical Telescope (UVOT)
aboard the {\it Swift Gamma-Ray Burst Mission}.  It is well-established that RR Lyrae have large
amplitudes in the far- and near-ultraviolet.  We have used UVOT's unique wide-field NUV imaging capability to perform the first 
systematic NUV survey of variable stars in the Galactic globular clusters M~3 and M~15.  We identify 280 variable stars, comprising 275
RR Lyrae, two anomalous Cepheids, one classical Cepheid, one SX Phoenicis star and one possible long-period or irregular variable.  Only
two of these are new discoveries.  We compare
our results to previous investigations and find excellent agreement in the periods with significantly larger amplitudes in the NUV.  We map out,
for the first time, an NUV Bailey diagram from globular clusters, showing the usual loci for fundamental mode RRab and first overtone
RRc pulsators.  We show the unique sensitivity of NUV photometry to both the temperatures and the surface gravities of RR Lyrae stars.  Finally, we
show evidence of an NUV period-metallicity-luminosity relationship.  Future investigations
will further examine the dependence of NUV pulsation parameters on metallicity and Oosterhoff classification.
\end{abstract}


\section{Introduction}
\label{s:intro}

RR Lyrae stars are moderate-mass horizontal branch (HB) stars with temperatures in the range of 6000-7600 K.  These ``cluster variables" have been studied for
over a century and become a key part of both the cosmic distance ladder and our understanding of the astrophysics of pulsating stars (see review in Smith 1995).
Tens of thousands of RR Lyrae have been identified in the Milky Way, of which several thousand are in old 
globular clusters (Clement et al. 2001, hereafter C01).

Since the early studies of Hardie (1955) and Baker \& Baker (1956), it has been known that RR Lyrae stars have exceptionally large pulsations
in the ultraviolet (UV), with amplitudes of 2-8 magnitudes (Hutchinson et al. 1977; Bonnell et al. 1982; Guhathakurta et al. 1994; Downes et al. 2004; Wheatley et al. 2005, 2012; Dieball et al. 2007;
Thomson et al. 2010).
These exceptionally large pulsations are due to the RR Lyrae occupying a unique niche in temperature-surface gravity space, where their oscillations
cause the UV flux to wax and wane dramatically.
These large pulsations could allow unique insight into the underlying astrophysics.
For example, the predicted UV fluxes of RR Lyrae stars at maximum and minimum light are very sensitive to
the particulars of underlying atmospheric models (see, e.g., Wheatley et al. 2005), with temperature sensitivity at the 100-200 K level.
UV light also probes different regions of stellar atmospheres than optical
light, allowing UV light curves to provide important constraint on hydrodynamical models of stellar pulsation, including the effect of shocks on
the outer stellar envelope (see, e.g., Hutchinson et al. 1977; Bonnell et al. 1982).

UV light curves could also improve the utility of RR Lyrae stars as distance indicators.  Estimating distances to stellar populations
using RR Lyrae stars is subject to a number 
of uncertainties that limit the precision of the method to about $\pm0.1$ mag.
However, the NIR passbands bypass the issues that complicate distance measures and provide more reliable and precise
RR Lyrae-based distances (see, e.g., Bono et al., 2001;  Del Principe et al. 2005; Sollima  et al. 2006; Neeley et al. 2015).  The horizontal branch in the NIR is diagonal, 
producing a direct relationship between period, luminosity and metallicity.  The horizontal branch is also diagonal in the NUV and FUV
(Ferraro \& Paresce, 1993, Dieball et al. 2007; Sandquist et al. 2010; Schiavon et al. 2012, Siegel et al. 2014, hereafter Paper I).
A correlation between NUV color and magnitude indicates a correlation between stellar temperature and magnitude and, thus, a likely
correlation between pulsational period and NUV magnitude similar to the one identified in the NIR.   If such a relationship were proven, a metallicity and pulsation period for an RR Lyrae
would yield its absolute magnitude {independent of reddening.  This would make RR Lyrae stars potent probes of local stellar populations
and reddening, the latter of which is highly uncertain in the UV (see, e.g., Pei 1992).

The most thorough surveys in the UV are
those of Downes et al. (2004), who presented partial light curves of 11 RR Lyrae in NGC~1852 using {\it Hubble Space Telescope} data and 
Wheatley et al. (2012), who presented detailed light curves for seven field RR Lyrae using GALEX data.  Both studies show
remarkable high-amplitude pulsations of up to eight magnitudes in the FUV.  To date, however, no study has presented detailed NUV or FUV
light curves of RR Lyrae stars in a globular cluster.
We now fill this gap with a survey of two Galactic globular clusters -- M~3 and M~15 -- that have rich populations of well-studied RR Lyrae stars
and represent the two poles of Oosterhoff (1939) dichotomy.

M~3 (NGC~5272) has the most well-studied population of variable stars of any Galactic globular cluster.  Indeed, one of the earliest indications that RR Lyrae stars
have large UV pulsations was from the study of M~3 by Baker \& Baker (1956).  C01 list 290 variable stars in M~3, 236 of which are RR Lyrae.
These stars have been the target of decades worth of study, most recently by
Kaluzny et al. (1998), Carretta et al. (1998), Bakos et al. (2000), Corwin \& Carney (2001),
Hartmann et al. (2005), Cacciari et al. (2005; hereafter CCC05) and Benko et al. (2006).  These studies have detailed M~3's abundant family of RR Lyrae stars and shown it to be a paradigm -- indeed
{\it the} paradigm -- of an Oosterhoff I (OoI) cluster, with a lower proportion of first-overtone RRc stars, shorter fundamental mode RRab periods and a shorter
minimum RRab period.  M~3 is often used as the baseline with which to measure the period-shift effect -- the increase or decrease in the period of RR Lyrae stars in comparison
to stars of equal pulsational amplitude (see Sandage 1981a, 1981b; Carney et al. 1992)

M~15 (NGC~7078) has also been studied extensively, as detailed in C01.  Studies by
Rosino (1950, 1969), Filippenko \& Simon (1981), Sandage et al. (1981), Bingham (1984), Silberman \& Smith (1995), Butler et al. (1998)
Tuairisg et al. (2003), Zheleznyak \& Kravtsov (2003) and Corwin et al (2008, hereafter C08) have revealed a rich population
of RR Lyrae stars that are a paradigm of a metal-poor Oosterhoff II (OoII) cluster, with a high proportion of RRc stars, longer RRab periods and a longer
minimum RRab period.

While both these clusters have been studied exhaustively in the optical and infrared, neither of these clusters has been thoroughly studied in the UV.
In this paper, we present the first thorough NUV catalog of the variable stars of these two clusters using data from the {\it Swift Gammy-Ray Burst Mission}'s 
Ultraviolet-Optical Telescope (UVOT).  Almost all of the variables studied are RR Lyrae stars, with a handful of other variable types also detected.  Our
analysis will therefore be focused on the RR Lyrae, investigating basic aspects of their pulsational properties and laying the groundwork for future investigations
of additional clusters.

\section{Observations and Data Reductions}
\label{s:obsred}

The data presented in this analysis are entirely from the UVOT aboard the Swift mission (Gehrels et al., 2004).  
UVOT is a modified Richey-Chretien 30 cm telescope that has a 
wide (17' $\times$ 17') field of view and a microchannel plate intensified CCD operating in photon
counting mode (see details in Roming et al. 2000, 2004, 2005).
The instrument is equipped with a filter wheel that includes a clear white filter, $u$, $b$ and $v$ optical filters, 
$uvw1$, $uvm2$ and $uvw2$ ultraviolet filters, a magnifier, two grisms and a blocked filter.  Although its primary
mission is to measure the optical/ultraviolet afterglows
of gamma ray bursts, the wide field, 2\farcs3 resolution, broad wavelength range (1700-8000 \AA)
and ability to observe simultaneously with {\it Swift}'s X-Ray Telescope (XRT; Burrows et al. 2005)
allow a broad range of science, including the study of hot or highly energetic stars (Paper I).  It is well-suited
to the study of star clusters due to its field size and resolution.

Our light curves are derived from observations taken with UVOT's $uvm2$ filter, which has a central wavelength of 2246 \AA\
and a red leak from optical light of less than 0.19\% (Paper I).  This was chosen to minimize potential contamination from the optical
to produce as clean an NUV light curve as possible (the $uvw1$ and $uvw2$ filters have optical sensitivities of 11\% and 2.4\%, respectively).
M~3 was observed on August 2011 as part of a Swift team fill-in program.  It was then observed 11 March 2012
through 6 June 2012 approximately every 4-8 days as part of MHS's team fill-in program, then again
from 4 April 2013 to 29 March 2014 as part of MHS's approved Guest Investigator program. M~15 was observed in May
and August of 2011 as a Target Of Opportunity, every four days from April 2013 to March of 2014
as part of MHS's approved Guest Investigator program and then again in October of 2014 as a team fill-in.
Table \ref{t:swiftphot} contains a summary of the observations.  

\begin{deluxetable}{c|c|ccc|c}
\tablewidth{0 pt}
\tabletypesize{\footnotesize}
\tablecaption{Swift/UVOT Observations of Globular Clusters\label{t:swiftphot}}
\tablehead{
\colhead{Cluster} &
\colhead{N Obs} &
\colhead{$uvw1$ Exp} &
\colhead{$uvm2$ Exp} &
\colhead{$uvw2$ Exp} &
\colhead{Observation}\\
\colhead{} &
\colhead{} &
\colhead{Time (ks)} &
\colhead{Time (ks)} &
\colhead{Time (ks)} &
\colhead{Dates}}
\startdata
\hline
M~3     & 67      & 1359    &  61400  & 1933    & 2011-08-31\\
        & \nodata & \nodata & \nodata & \nodata & 2012-03-11 -- 2012-06-12\\
        & \nodata & \nodata & \nodata & \nodata & 2013-04-04 -- 2014-03-29\\
\hline
M~15    & 55      & 5203    &  44849  &  287    & 2011-05-18 -- 2011-08-28\\
        & \nodata & \nodata & \nodata & \nodata & 2013-04-11 -- 2014-03-31\\
        & \nodata & \nodata & \nodata & \nodata & 2014-10-16\\
\hline
\enddata
\end{deluxetable}

For the monitoring program, each cluster was scheduled for a nominal 1 ks for each observation.  However, Swift's planning sometimes broke that observation
up over two or more orbits, separated by some multiple of Swift's 96-minute orbital period.  Observations that were not
performed as part of the monitoring campaign were sometimes performed in modes that did not include $uvm2$.
Additionally, some observations were curtailed or cancelled by gamma-ray bursts or high-priority targets or had trailed images.
In the end, our analysis of M~3 is based on 85 individual $uvm2$ images and M~15 from 48 individual $uvm2$ images.

Photometry was derived using DAOPHOT/ALLFRAME (Stetson 1987, 1994).  The details of this process can be found in Paper I.  To briefly summarize: 
we created individual images for each observation from the archival data downloaded from HEASARC, regenerating the exposure maps and large-scale
sensitivity images for calibration.  After running DAOPHOT on the individual fields, the images were then put through
ALLFRAME to derive consistent deep photometry on all fields.  We then calibrated the instrumental photometry
to the standard system (Poole et al. 2008; Breeveld at al. 2010, 2011) correcting for coincidence loss, variation in exposure time and large-scale sensitivity.  

\begin{figure}[h]
\begin{center}
\includegraphics[scale=1.]{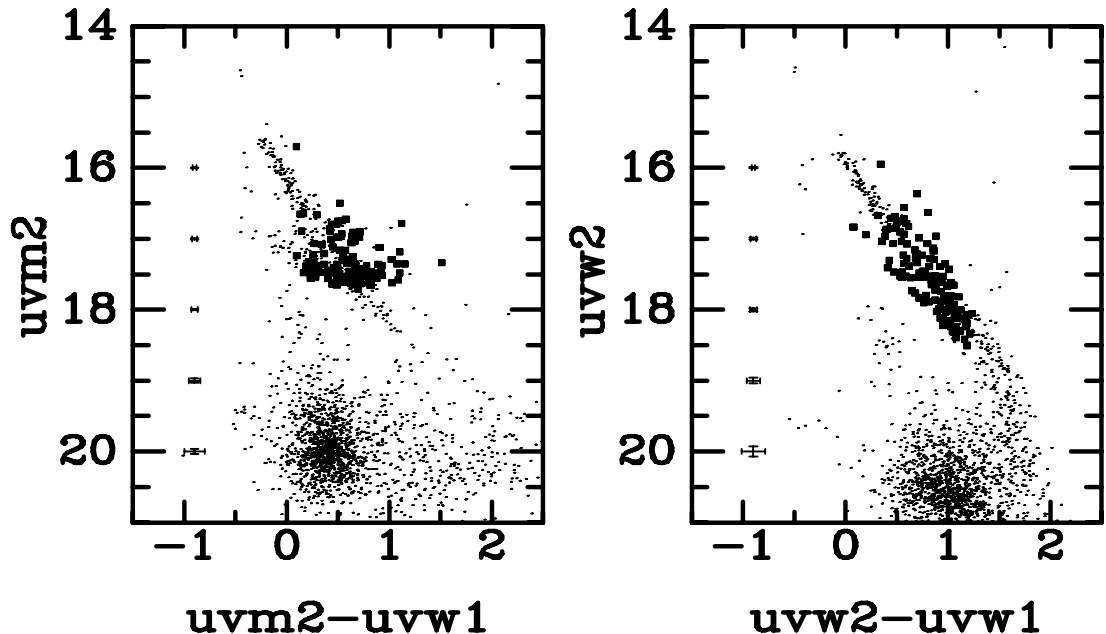}
\end{center}
\caption{UVOT color-magnitude diagrams of the globular cluster M~3.  The left panel plots $uvm2$ against $uvw1$ while the right
plot $uvw2$ against $uvw1$.
The prominent diagonal sequence is the HB.  Blue stragglers, AGB manque stars and other hot star types are present, as detailed
in Paper I.  The squares mark RR Lyrae identified in this study.  Error bars on the left side indicate typical uncertainties.\label{f:m3cmd}}
\end{figure}

\begin{figure}[h]
\begin{center}
\includegraphics[scale=1]{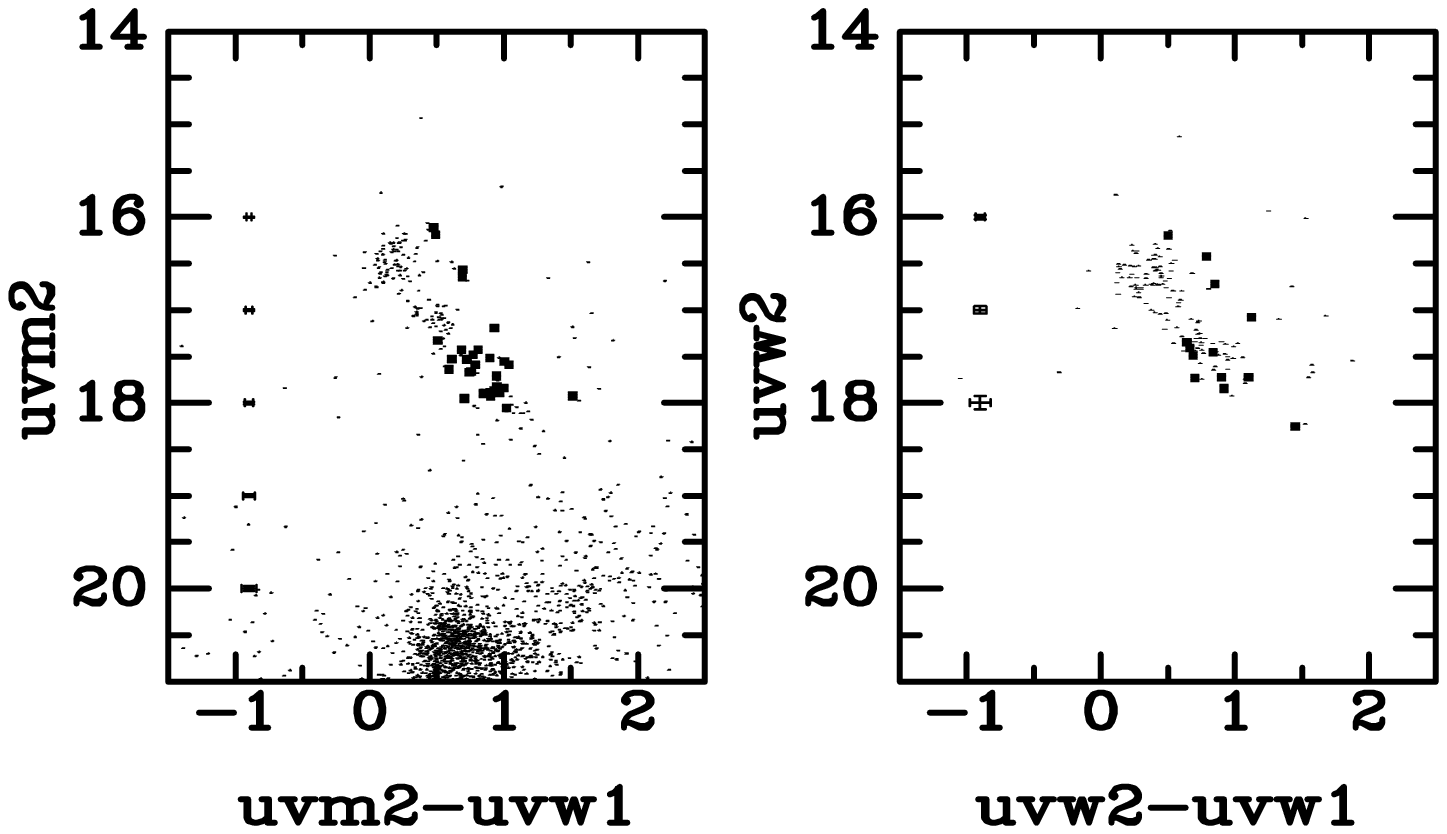}
\end{center}
\caption{UVOT color-magnitude diagrams of the globular cluster M~15.  The left panel plots $uvm2$ against $uvw1$ while the right
plot $uvw2$ against $uvw1$.
The prominent diagonal sequence
is the HB.  Blue stragglers, AGB manque stars and other hot star types are present, as detailed
in Paper I.  The $uvw2$ data are very shallow due to the very small amount of time invested
in the filter during the observational campaign.  The squares mark RR Lyrae identified in this study.  Error bars on the left side indicate typical uncertainties.\label{f:m15cmd}}
\end{figure}

Figures \ref{f:m3cmd} and \ref{f:m15cmd} show the color-magnitude diagrams of M~3 and M~15, respectively.  We should $uvm2-uvw1$ on the left and $uvw2-uvw1$ on the right.
In both cases, we have only plotted stars with DAOPHOT
SHARP parameters of 0.5 or less.  SHARP is a structural parameter produced by DAOPHOT that measures the difference between the isophotal and peak magnitudes.
We have found it to be the most effective means of identifying poor measures in UVOT data and, in these cases, it mostly excludes stars from the
burned-in cores of the clusters.

The most visible features are the diagonal horizontal branches,
with the RR Lyrae gap visible in the increased scatter around the HB locus.  The faint haze of stars at the bottom of the CMD
are mostly bright RGB and field stars detected by the red leak.

\subsection{Light Curve Fitting}
\label{s:lcf}

Variable stars were selected for light curve fitting using the variability index measured by the DAOMASTER code included with DAOPHOT.  Any
star with a photometric scatter more than three times the formal error was selected as a potential variable star.
We further refined the list by only selecting stars with DAOPHOT SHARP values less than 0.5.  This excluded many stars that were within
the burned-in core of the cluster.

Our first pass at the light curves used the RRFIT code of Yang \& Sarajedini (2012), which fits the light curves
using the light curve templates of Layden (1998) as well as those of Kovacs \& Kupi (2007).  These templates are drawn from a sample of RR Lyrae
stars and represent a wide variety of light curve shapes.  The drawback of template-fitting is that
it can be confused by second order variations such as the Blazhko (1907) effect or double-mode pulsation.  It also may not be entirely appropriate to
NUV data, which may have different light curve shapes than the optical data on which the templates are based.  Finally, our search parameters only
covered periods between 0.2 to 1.2 days, which would exclude most SX Phoenicis stars, Cepheids and long-period variables.  Nevertheless, this provided a good first estimate
of the periods and amplitudes for the vast majority of the variable stars in these two systems.

After preliminary fits were obtained with RRFIT, we ran each of our stars through IRAF's phase-dispersion minimization algorithm to both refine
the period and provide an eyeball check on the fits.  For almost all stars, the two methods agreed.  For those where there
was some disagreement due to multiple minima in the periodogram, we used the period with the lowest phase dispersion.  For a handful of
stars, neither method produced a satisfactory result.

After fitting the light curves based on UVOT data, we compared our results to the comprehensive C01 catalog of globular cluster variable
stars.  We pulled out for analysis any variable star candidates from C01 that had not been selected as variable star candidates
from the UVOT data (either because their variability
index was too low or their SHARP parameter too high)  For most of these, we were able to
fit periods and amplitudes using the UVOT data, both with template-fitting and with phase dispersion minimization.

Tables \ref{t:M3cat} and \ref{t:M15cat} provide the catalog of variable stars, with identification numbers taken from C01.  Where we
have identified new variable stars, we have added new identification numbers sequential to C01.
The columns list C01 identification number,
equatorial coordinates, periods, amplitudes and mean magnitudes derived from the $UVOT$ data.
The last two columns list the variability index and the SHARP parameter as calculated by DAOPHOT.
Tables \ref{t:M3phot} and \ref{t:M15phot} give the Julian Dates, $uvm2$ measures and measurement uncertainties for the variable stars.

Figure \ref{f:M3RRL0} shows the light curves for all of the RR Lyrae stars in M~3, 
ordered by increasing period.  Figure \ref{f:M3other} shows the light curves of the other variable stars measured in M~3 and Figure \ref{f:M3long}
shows the light curve of a possible long-period or irregular variable.
Figures \ref{f:M15RRL0} shows the light curves for all of the RR Lyrae stars in M~15, 
ordered by increasing period.  Figure \ref{f:M15other} shows the light curves of the other variable stars measured in M~15.

One of the pitfalls of using previous catalogs to pull marginal detections out of the noise is that the light curves of such stars are less than ideal.  UVOT's 
comparatively coarse resolution means that stars in the cores of globular clusters frequently have their light blended with that of nearby stars.  This tends to increase
the magnitudes and dampen the amplitudes of any pulsations, muddying the correlations between pulsation parameters (see \S \ref{s:analysis}).  This can be seen visually
in Figures \ref{f:M3RRL0} through \ref{f:M15other}, where we have marked with asterisks stars with DAOPHOT SHARP values greater than 0.5 -- which indicates a broad PSF.
These stars are found almost exclusively in the burned-in cores of the clusters.  

The effect of blends on the UVOT photometry is complicated, raising the measured count rate but also increasing the coincidence loss.  It is impossible to remove
their contribution without making {\it a priori} assumptions about the variable stars or their unresolved companions.
We include these stars in the compilation but caution that their $uvm2$ 
magnitudes and $A_{uvm2}$ amplitudes are {\bf not reliable} and should be excluded from any global analysis of the variable stars.

For completeness, our tables include mean magnitudes, variability and SHARP values for stars which  we have matched to the C01 catalog but are unable to fit a period.
This includes stars that C01 classify as non-variable.  However, these stars are only included if the SHARP value is less than 0.5.  Stars with SHARP values greater
than 0.5 may be blends and are much more likely to be have been misidentified or have their pulsations drowned in the light of nearby star.

\begin{deluxetable}{lccccccc}
\tablewidth{0 pt}
\tabletypesize{\footnotesize}
\tablecaption{Parameters of Variable Stars in M~3\label{t:M3cat}}
\tablehead{
\colhead{ID} &
\colhead{$\alpha$} &
\colhead{$\delta$} &
\colhead{P} &
\colhead{$<uvm2>_{i}$} &
\colhead{$A_{uvm2}$} &
\colhead{Var} &
\colhead{SHARP} \\
\colhead{(C01)} &
\multicolumn{2}{c}{(J2000.0)} &
\colhead{(days)} &
\colhead{} &
\colhead{} &
\colhead{} &
\colhead{}}
\startdata
\hline
        V1  &    13:42:11.11  &     28:20:33.5  &     0.5205563  &   17.338  &    2.382  &     6.60  &     0.15   \\
        V3  &    13:42:15.70  &     28:21:41.0  &     0.5581785  &   17.289  &    2.438  &     4.19  &     0.35   \\
       V4*  &    13:42:08.19  &     28:22:32.9  &     0.5850817  &   16.395  &    1.125  &     5.22  &     0.66   \\
        V5  &    13:42:31.26  &     28:22:20.4  &     0.5058278  &   17.116  &    1.650  &     6.43  &     0.28   \\
        V6  &    13:42:02.06  &     28:23:41.3  &     0.5143057  &   17.516  &    2.320  &     4.66  &     0.27   \\
        V7  &    13:42:11.08  &     28:24:09.9  &     0.4974221  &   17.305  &    2.341  &     5.74  &     0.49   \\
        V8  &    13:42:05.32  &     28:22:18.4  &     0.6367176  &   16.784  &    0.709  &     5.20  &     0.29   \\
        V9  &    13:41:49.50  &     28:19:13.3  &     0.5415324  &   17.463  &    2.316  &     4.54  &     0.14   \\
       V10  &    13:42:23.08  &     28:25:00.2  &     0.5695330  &   17.459  &    1.764  &     5.32  &     0.19   \\
       V11  &    13:41:59.97  &     28:19:11.7  &     0.5078967  &   17.239  &    2.181  &     5.81  &     0.35   \\
       V12  &    13:42:11.25  &     28:20:16.8  &     0.3179041  &   16.750  &    0.898  &     6.33  &     0.41   \\
       V13  &    13:42:09.57  &     28:20:24.3  &     0.4795181  &   17.163  &    1.183  &     6.62  &     0.19   \\
       V14  &    13:42:07.79  &     28:20:00.9  &     0.6358649  &   17.546  &    2.131  &     6.98  &     0.12   \\
       V15  &    13:42:04.66  &     28:18:08.5  &     0.5300817  &   17.385  &    2.268  &     5.92  &     0.21   \\
       V16  &    13:41:48.70  &     28:21:07.9  &     0.5114729  &   17.507  &    2.312  &     4.31  &     0.26   \\
       V17  &    13:42:22.40  &     28:15:22.6  &     0.5761198  &   17.528  &    1.401  &     6.32  &     0.14   \\
       V18  &    13:42:18.94  &     28:17:47.0  &     0.5164755  &   17.597  &    2.341  &     5.34  &     0.10   \\
       V19  &    13:42:38.10  &     28:18:37.5  &     0.6320068  &   17.709  &    1.207  &     4.18  &     0.14   \\
       V20  &    13:42:36.81  &     28:18:11.5  &     0.4912951  &   17.185  &    1.924  &     6.42  &     0.17   \\
       V21  &    13:42:37.76  &     28:23:01.3  &     0.5157982  &   17.558  &    2.406  &     5.01  &     0.15   \\
       V22  &    13:42:25.89  &     28:22:31.6  &     0.4813960  &   17.298  &    2.212  &     6.05  &     0.12   \\
       V23  &    13:42:02.83  &     28:27:20.1  &     0.5954373  &   17.465  &    1.462  &     5.16  &     0.21   \\
       V24  &    13:42:00.30  &     28:22:51.6  &     0.6634306  &   17.460  &    1.799  &     5.01  &     0.26   \\
       V25  &    13:42:02.07  &     28:22:09.9  &     0.4800817  &   17.372  &    2.316  &     3.71  &     0.13   \\
       V26  &    13:41:58.03  &     28:21:58.2  &     0.5977560  &   17.328  &    1.948  &     5.31  &     0.28   \\
       V27  &    13:42:03.16  &     28:20:58.8  &     0.5790610  &   17.579  &    1.801  &     5.46  &    -0.20   \\
       V28  &    13:42:09.61  &     28:20:56.2  &     0.4706749  &   17.122  &    1.338  &     6.63  &     0.18   \\
      V29*  &    13:42:06.64  &     28:21:27.4  &     0.4717771  &   16.318  &    0.690  &     5.03  &     0.84   \\
       V30  &    13:42:08.73  &     28:23:39.6  &     0.5120514  &   17.373  &    2.284  &     5.97  &     0.15   \\
       V31  &    13:42:13.99  &     28:23:47.2  &     0.5807016  &   17.661  &    2.363  &     5.88  &    -0.25   \\
      V32*  &    13:42:12.38  &     28:23:41.9  &     0.4953458  &   17.054  &    1.821  &     4.92  &     0.51   \\
       V33  &    13:42:16.82  &     28:21:13.0  &     0.5252704  &   17.349  &    1.808  &     5.30  &    -0.15   \\
       V34  &    13:42:21.70  &     28:25:32.2  &     0.5591055  &   17.445  &    1.471  &     7.37  &     0.10   \\
       V35  &    13:42:03.43  &     28:18:03.4  &     0.5305683  &   17.362  &    1.713  &     6.21  &     0.21   \\
       V36  &    13:42:24.52  &     28:22:07.1  &     0.5455711  &   17.509  &    2.158  &     5.20  &     0.09   \\
       V37  &    13:41:53.55  &     28:25:25.2  &     0.3266509  &   16.931  &    1.015  &     7.29  &     0.19   \\
       V38  &    13:41:56.04  &     28:24:48.6  &     0.5580155  &   17.480  &    1.604  &     6.91  &     0.10   \\
       V39  &    13:41:52.96  &     28:24:42.0  &     0.5870736  &   17.597  &    1.472  &     6.17  &     0.04   \\
       V40  &    13:41:50.93  &     28:24:32.6  &     0.5515324  &   17.611  &    2.100  &     4.57  &     0.13   \\
       V41  &    13:42:04.37  &     28:23:35.5  &     0.4850817  &   17.447  &    2.418  &     5.23  &    -0.20   \\
       V42  &    13:42:05.50  &     28:23:22.0  &     0.5900818  &   17.225  &    2.062  &     5.90  &     0.26   \\
       V43  &    13:42:19.05  &     28:23:06.6  &     0.5404946  &   17.545  &    2.084  &     5.40  &    -0.18   \\
       V44  &    13:42:24.35  &     28:24:21.7  &     0.5065204  &   17.219  &    1.450  &     7.00  &     0.16   \\
       V45  &    13:41:53.22  &     28:20:31.1  &     0.5368882  &   17.548  &    1.992  &     6.40  &     0.10   \\
       V46  &    13:42:01.80  &     28:21:50.2  &     0.6133907  &   17.559  &    1.218  &     4.67  &     0.43   \\
       V47  &    13:42:02.59  &     28:21:28.5  &     0.5410013  &   17.356  &    1.666  &     6.87  &     0.16   \\
       V48  &    13:42:21.15  &     28:21:00.1  &     0.6279041  &   17.380  &    1.340  &     5.08  &     0.33   \\
       V49  &    13:42:22.09  &     28:21:02.4  &     0.5482007  &   17.463  &    1.716  &     7.85  &     0.34   \\
       V50  &    13:42:12.22  &     28:18:47.8  &     0.5131192  &   17.373  &    1.971  &     6.66  &     0.05   \\
       V51  &    13:42:13.85  &     28:18:55.8  &     0.5840006  &   17.583  &    1.688  &     4.65  &     0.11   \\
      V52*  &    13:42:05.62  &     28:25:12.8  &     0.5162355  &   17.212  &    1.633  &     5.57  &     0.90   \\
       V53  &    13:42:10.92  &     28:24:44.5  &     0.5048887  &   17.356  &    2.084  &     4.67  &     0.17   \\
       V54  &    13:42:08.98  &     28:24:28.2  &     0.5061939  &   17.209  &    1.366  &     5.93  &     0.11   \\
       V55  &    13:41:55.91  &     28:28:05.3  &     0.5298442  &   17.517  &    2.188  &     4.36  &     0.10   \\
       V56  &    13:42:00.67  &     28:28:39.4  &     0.3296071  &   16.892  &    0.926  &     6.74  &     0.12   \\
       V57  &    13:42:23.23  &     28:22:41.9  &     0.5121922  &   17.493  &    2.354  &     3.95  &    -0.01   \\
      V58*  &    13:42:04.94  &     28:23:27.6  &     0.5170514  &   17.169  &    2.125  &     5.60  &     0.56   \\
       V59  &    13:42:03.23  &     28:18:53.1  &     0.5887867  &   17.636  &    1.779  &     5.81  &     0.17   \\
       V60  &    13:41:49.05  &     28:17:25.5  &     0.7077409  &   17.592  &    1.729  &     5.08  &     0.21   \\
       V61  &    13:42:25.77  &     28:28:45.3  &     0.5209119  &   17.366  &    1.649  &     7.87  &     0.14   \\
       V62  &    13:42:18.21  &     28:29:38.7  &     0.6524221  &   17.639  &    1.440  &     5.11  &     0.26   \\
       V63  &    13:42:14.22  &     28:28:23.6  &     0.5703336  &   17.557  &    1.962  &     6.17  &     0.17   \\
       V64  &    13:42:20.10  &     28:28:12.0  &     0.6054125  &   17.623  &    1.476  &     5.00  &     0.27   \\
       V65  &    13:42:20.90  &     28:28:09.6  &     0.6683648  &   17.473  &    1.994  &    14.60  &     0.26   \\
       V66  &    13:42:03.76  &     28:24:42.4  &     0.6201262  &   17.553  &    1.256  &     6.18  &     0.35   \\
       V67  &    13:42:01.50  &     28:24:44.0  &     0.5683120  &   17.499  &    1.720  &     4.24  &     0.46   \\
       V68  &    13:42:13.08  &     28:25:36.7  &     0.3559851  &   16.977  &    1.104  &     7.49  &     0.34   \\
       V69  &    13:42:17.56  &     28:25:03.0  &     0.5666287  &   17.607  &    1.999  &     5.51  &     0.09   \\
       V70  &    13:42:14.32  &     28:25:14.0  &     0.4861218  &   16.979  &    0.694  &     6.66  &     0.11   \\
       V71  &    13:42:23.63  &     28:22:40.0  &     0.5490881  &   17.505  &    1.718  &     6.24  &     0.13   \\
       V72  &    13:42:45.23  &     28:22:40.9  &     0.4561051  &   17.360  &    2.436  &     3.82  &     0.19   \\
       V73  &    13:42:44.70  &     28:23:45.4  &     0.6734648  &   17.624  &    0.647  &     4.73  &     0.29   \\
       V74  &    13:42:18.13  &     28:25:12.7  &     0.4921432  &   17.427  &    2.333  &     5.41  &     0.24   \\
       V75  &    13:42:15.12  &     28:25:21.0  &     0.3140880  &   16.903  &    0.928  &     7.56  &     0.21   \\
       V76  &    13:42:10.40  &     28:21:13.8  &     0.5017919  &   17.539  &    2.549  &     5.44  &    -0.16   \\
       V77  &    13:42:04.28  &     28:23:09.3  &     0.4593696  &   17.472  &    2.717  &     5.10  &    -0.02   \\
       V78  &    13:42:15.05  &     28:23:48.3  &     0.6118882  &   17.406  &    1.340  &     5.43  &    -0.00   \\
       V79  &    13:42:14.70  &     28:28:31.2  &     0.4833121  &   17.373  &    2.432  &     5.75  &     0.23   \\
       V80  &    13:42:43.02  &     28:27:27.5  &     0.5384485  &   17.434  &    2.130  &     6.30  &     0.01   \\
       V81  &    13:42:37.35  &     28:28:33.9  &     0.5291203  &   17.464  &    2.254  &     4.06  &     0.18   \\
       V83  &    13:41:38.00  &     28:24:33.2  &     0.5012950  &   17.477  &    2.403  &     5.56  &     0.10   \\
       V84  &    13:42:16.30  &     28:25:26.8  &     0.5957091  &   17.560  &    1.587  &     6.76  &     0.24   \\
       V85  &    13:42:34.64  &     28:26:28.7  &     0.3558278  &   16.797  &    1.083  &    10.78  &     0.18   \\
       V86  &    13:42:50.39  &     28:20:48.7  &     0.2926594  &   16.909  &    1.079  &     5.49  &     0.21   \\
       V87  &    13:42:19.83  &     28:23:42.2  &     0.3574923  &   16.852  &    0.711  &     5.35  &     0.37   \\
       V88  &    13:42:08.82  &     28:21:32.0  &     0.2987496  &   16.964  &    1.239  &     6.60  &     0.09   \\
       V89  &    13:42:13.61  &     28:20:51.3  &     0.5484975  &   17.650  &    2.447  &     6.82  &    -0.16   \\
       V90  &    13:42:18.90  &     28:19:33.9  &     0.5170495  &   17.465  &    2.257  &     4.78  &     0.29   \\
      V91*  &    13:42:10.56  &     28:13:32.2  &     0.5301558  &   17.387  &    2.543  &    25.96  &     0.60   \\
       V92  &    13:42:09.37  &     28:15:53.4  &     0.5035271  &   17.392  &    2.131  &     6.17  &     0.12   \\
       V93  &    13:41:47.40  &     28:16:04.2  &     0.6023110  &   17.581  &    1.598  &     4.58  &     0.18   \\
       V94  &    13:41:34.55  &     28:18:55.2  &     0.5237049  &   17.552  &    1.940  &     3.54  &     0.15   \\
       V96  &    13:41:59.11  &     28:18:47.3  &     0.4993844  &   17.248  &    2.243  &     4.99  &     0.37   \\
       V97  &    13:42:01.69  &     28:19:24.6  &     0.3349331  &   17.080  &    0.960  &     6.82  &     0.12   \\
       V98  &    13:42:21.50  &     28:22:38.8  &       \nodata  &   16.418  &  \nodata  &     1.62  &     0.27   \\
       V99  &    13:42:26.73  &     28:21:47.6  &     0.4822200  &   17.008  &    0.509  &     7.09  &     0.19   \\
      V100  &    13:42:16.74  &     28:24:19.5  &     0.6187912  &   17.620  &    1.366  &     5.46  &     0.16   \\
      V101  &    13:42:14.98  &     28:24:05.3  &     0.6439074  &   17.635  &    1.266  &     5.07  &     0.36   \\
      V102  &    13:42:15.86  &     28:24:36.9  &       \nodata  &   17.896  &  \nodata  &     2.00  &     0.32   \\
      V104  &    13:42:09.49  &     28:25:07.0  &     0.5699183  &   17.379  &    2.359  &     5.87  &     0.26   \\
      V105  &    13:42:09.84  &     28:25:52.9  &     0.2877559  &   16.646  &    0.580  &     5.17  &     0.19   \\
      V106  &    13:42:07.79  &     28:25:29.6  &     0.5468882  &   17.558  &    1.846  &     5.80  &     0.48   \\
      V107  &    13:42:05.63  &     28:28:16.3  &     0.3090314  &   16.927  &    1.105  &     6.03  &     0.18   \\
      V108  &    13:41:54.78  &     28:27:51.6  &     0.5196071  &   17.533  &    2.294  &     5.46  &     0.15   \\
      V109  &    13:42:04.71  &     28:22:44.0  &     0.5338904  &   17.485  &    2.229  &     5.19  &     0.38   \\
      V110  &    13:42:03.93  &     28:22:25.8  &     0.5354818  &   17.460  &    1.944  &     6.10  &     0.01   \\
     V111*  &    13:42:04.44  &     28:23:03.5  &     0.5102003  &   17.249  &    1.585  &     4.78  &     0.77   \\
      V117  &    13:42:18.38  &     28:14:53.9  &     0.6005268  &   17.503  &    1.720  &     5.13  &     0.13   \\
      V118  &    13:42:22.49  &     28:17:50.3  &     0.4993696  &   17.473  &    2.100  &     5.81  &     0.19   \\
      V119  &    13:42:30.65  &     28:24:28.5  &     0.5176224  &   17.319  &    2.055  &     4.84  &     0.30   \\
      V120  &    13:41:48.98  &     28:26:31.7  &     0.6401262  &   17.572  &    1.048  &     4.49  &     0.40   \\
      V121  &    13:42:08.16  &     28:23:37.5  &     0.5352151  &   17.482  &    2.035  &     5.87  &     0.02   \\
      V124  &    13:42:06.50  &     28:19:20.5  &     0.7524591  &   17.614  &    0.789  &     5.19  &     0.24   \\
      V125  &    13:42:25.62  &     28:20:29.9  &     0.3498442  &   17.031  &    0.783  &     5.68  &     0.19   \\
      V126  &    13:42:10.34  &     28:20:15.5  &     0.3484204  &   17.071  &    0.825  &     7.53  &     0.09   \\
      V128  &    13:42:20.11  &     28:24:53.2  &     0.2920440  &   16.797  &    1.060  &     8.24  &     0.22   \\
      V129  &    13:42:08.20  &     28:23:59.1  &     0.4060800  &   16.976  &    0.841  &     5.07  &     0.14   \\
      V130  &    13:42:11.76  &     28:24:05.9  &     0.5689839  &   17.054  &    0.927  &     4.82  &     0.38   \\
      V131  &    13:42:05.88  &     28:23:08.7  &     0.2976816  &   16.906  &    0.912  &     4.86  &     0.09   \\
      V132  &    13:42:07.42  &     28:22:19.8  &     0.3398590  &   16.892  &    0.751  &     4.76  &     0.38   \\
      V133  &    13:42:07.00  &     28:23:25.3  &     0.5507461  &   17.443  &    2.064  &     4.39  &     0.25   \\
      V134  &    13:42:09.77  &     28:23:34.2  &     0.6180969  &   17.489  &    1.362  &     4.79  &     0.30   \\
     V135*  &    13:42:09.44  &     28:23:20.0  &     0.5683759  &   17.292  &    1.392  &     3.95  &     0.65   \\
     V137*  &    13:42:15.45  &     28:22:22.9  &     0.5751706  &   17.172  &    1.511  &     3.79  &     0.64   \\
      V138  &    13:41:51.52  &     28:23:22.3  &       \nodata  &   20.915  &  \nodata  &     1.10  &    -0.03   \\
     V139*  &    13:42:14.07  &     28:23:10.3  &     0.5599315  &   16.917  &    1.798  &     3.51  &     1.30   \\
      V140  &    13:42:10.28  &     28:24:30.4  &     0.3331340  &   16.744  &    0.855  &     6.71  &     0.05   \\
     V142*  &    13:42:09.24  &     28:21:43.6  &     0.5686457  &   17.732  &    2.222  &     5.47  &     0.51   \\
      V144  &    13:42:15.58  &     28:21:02.5  &     0.5968290  &   17.420  &    0.829  &     5.03  &     0.32   \\
     V145*  &    13:42:13.72  &     28:22:51.1  &     0.5144810  &   15.423  &    0.373  &     1.35  &     1.07   \\
      V146  &    13:42:18.48  &     28:21:43.6  &     0.5021922  &   17.353  &    2.313  &     5.10  &     0.48   \\
     V147*  &    13:42:09.83  &     28:23:29.1  &     0.3464951  &   16.814  &    0.628  &     4.24  &     0.71   \\
      V149  &    13:42:14.07  &     28:23:34.8  &     0.5482007  &   17.499  &    2.116  &     4.75  &     0.41   \\
      V150  &    13:42:16.67  &     28:23:20.0  &     0.5239690  &   17.446  &    2.009  &     4.64  &     0.30   \\
     V151*  &    13:42:12.01  &     28:22:01.3  &     0.5168290  &   16.809  &    1.150  &     3.54  &     0.90   \\
      V152  &    13:42:17.45  &     28:23:33.2  &     0.3261199  &   16.652  &    0.763  &     6.83  &     0.41   \\
     V154*  &    13:42:11.65  &     28:22:14.2  &    15.2841997  &   15.418  &    1.151  &     5.13  &     1.04   \\
     V156*  &    13:42:09.95  &     28:22:01.3  &     0.5319919  &   17.158  &    1.761  &     4.56  &     0.51   \\
     V157*  &    13:42:10.20  &     28:23:19.0  &     0.5428745  &   16.101  &    0.458  &     2.08  &     0.90   \\
     V160*  &    13:42:10.79  &     28:21:58.8  &     0.6573062  &   17.109  &    1.413  &     4.94  &     0.59   \\
     V161*  &    13:42:12.76  &     28:21:44.7  &     0.5265768  &   17.017  &    1.340  &     4.57  &     0.64   \\
     V162*  &    13:42:13.57  &     28:22:11.1  &     0.5448409  &   15.915  &    0.325  &     1.79  &     0.92   \\
      V165  &    13:42:17.03  &     28:23:02.6  &     0.4836457  &   17.271  &    2.462  &     4.89  &     0.41   \\
      V166  &    13:42:04.15  &     28:22:34.2  &     0.4850076  &   17.159  &    1.038  &     6.69  &     0.19   \\
      V167  &    13:42:05.58  &     28:22:05.4  &     0.6440240  &   17.545  &    0.960  &     4.09  &     0.04   \\
     V168*  &    13:42:08.06  &     28:22:49.4  &     0.2759464  &   16.502  &    0.541  &     4.05  &     0.67   \\
     V170*  &    13:42:09.35  &     28:23:14.9  &     0.4326001  &   16.453  &    0.686  &     3.19  &     1.04   \\
     V171*  &    13:42:09.51  &     28:22:58.7  &     0.3032750  &   16.572  &    0.803  &     3.02  &     1.00   \\
      V172  &    13:42:09.86  &     28:23:08.3  &     0.5422813  &   17.455  &    1.986  &     4.61  &     0.14   \\
     V174*  &    13:42:10.82  &     28:22:09.0  &     0.5913617  &   17.098  &    0.955  &     2.90  &     1.42   \\
     V175*  &    13:42:14.65  &     28:23:09.0  &     0.5696960  &   17.112  &    1.238  &     2.74  &     1.03   \\
     V176*  &    13:42:14.96  &     28:23:15.3  &     0.5396516  &   17.263  &    1.639  &     4.92  &     0.72   \\
      V177  &    13:42:16.24  &     28:22:13.5  &     0.3483463  &   16.723  &    0.878  &     6.74  &     0.27   \\
     V178*  &    13:42:17.45  &     28:23:29.1  &     0.2669697  &   16.663  &    0.601  &     4.93  &     0.63   \\
     V180*  &    13:42:10.16  &     28:22:13.9  &     0.6091277  &   16.959  &    0.580  &     1.86  &     1.55   \\
      V182  &    13:42:10.10  &     28:23:41.5  &       \nodata  &   18.020  &  \nodata  &     2.13  &     0.37   \\
     V184*  &    13:42:09.58  &     28:22:27.7  &     0.5311763  &   17.086  &    1.624  &     2.72  &     1.61   \\
     V186*  &    13:42:12.47  &     28:21:39.4  &     0.6631933  &   17.327  &    0.911  &     4.55  &     0.80   \\
     V190*  &    13:42:10.89  &     28:23:10.7  &     0.5228004  &   17.019  &    1.340  &     3.11  &     1.10   \\
     V191*  &    13:42:11.57  &     28:23:05.9  &     0.5191869  &   16.262  &    0.697  &     2.53  &     0.56   \\
     V192*  &    13:42:11.32  &     28:22:46.9  &     0.4973590  &   15.375  &    0.410  &     2.32  &     1.14   \\
     V195*  &    13:42:10.47  &     28:22:14.2  &     0.6439052  &   16.937  &    0.521  &     2.68  &     1.35   \\
      V197  &    13:42:15.88  &     28:22:51.4  &     0.4999183  &   17.476  &    2.667  &     6.61  &    -0.21   \\
     V200*  &    13:42:11.16  &     28:23:03.8  &     0.3609884  &   16.717  &    0.761  &     3.59  &     0.79   \\
     V201*  &    13:42:11.81  &     28:22:33.8  &     0.5405832  &   15.611  &    0.338  &     1.94  &     0.95   \\
      V202  &    13:41:42.83  &     28:24:17.2  &     0.7735864  &   17.579  &    0.455  &     4.01  &     0.19   \\
      V203  &    13:42:09.43  &     28:17:30.2  &     0.2897997  &   16.661  &    0.277  &     2.98  &     0.22   \\
      V204  &    13:42:38.69  &     28:22:47.3  &       \nodata  &   17.527  &  \nodata  &     1.52  &     0.21   \\
      V207  &    13:42:14.16  &     28:22:11.4  &     0.3453130  &   16.500  &    0.478  &     3.11  &     0.41   \\
     V208*  &    13:42:11.59  &     28:21:44.3  &     0.3696448  &   15.606  &    0.212  &     2.03  &     0.72   \\
      V209  &    13:42:06.37  &     28:21:02.9  &     0.3483121  &   15.699  &    0.285  &     2.69  &     0.31   \\
     V210*  &    13:42:04.89  &     28:22:31.2  &     0.3529560  &   15.945  &    0.264  &     2.18  &     0.52   \\
     V211*  &    13:42:07.42  &     28:22:50.7  &     0.5581773  &   16.235  &    0.362  &     1.59  &     0.84   \\
     V212*  &    13:42:09.86  &     28:22:04.6  &     0.5421922  &   17.194  &    1.087  &     3.25  &     0.99   \\
     V213*  &    13:42:09.57  &     28:22:12.3  &     0.3000373  &   16.458  &    0.726  &     2.91  &     0.99   \\
     V215*  &    13:42:10.45  &     28:22:41.0  &     0.5287049  &   16.491  &    1.257  &     2.61  &     1.65   \\
     V216*  &    13:42:13.58  &     28:22:32.1  &     0.3464803  &   16.375  &    0.479  &     2.40  &     1.66   \\
     V217*  &    13:42:11.37  &     28:22:14.9  &     0.5283713  &   15.771  &    0.453  &     2.74  &     0.74   \\
     V219*  &    13:42:07.04  &     28:22:57.4  &     0.6136531  &   17.346  &    1.208  &     5.50  &     0.52   \\
     V220*  &    13:42:13.99  &     28:22:26.2  &     0.6001558  &   17.093  &    1.043  &     3.35  &     0.70   \\
      V222  &    13:42:18.75  &     28:21:38.6  &     0.5967771  &   17.292  &    1.513  &     4.13  &     0.39   \\
     V223*  &    13:42:13.24  &     28:22:36.2  &     0.3292243  &   16.381  &    0.651  &     2.69  &     1.42   \\
     V229*  &    13:42:08.98  &     28:21:55.7  &     0.4976213  &   16.471  &    1.288  &     2.98  &     0.57   \\
      V230  &    13:42:30.92  &     28:14:29.1  &    20.2900009  &    0.000  &    1.850  &     0.22  &     0.00   \\
      V231  &    13:42:20.91  &     28:23:29.2  &       \nodata  &   16.043  &  \nodata  &     1.22  &     0.17   \\
      V232  &    13:42:20.49  &     28:23:23.7  &       \nodata  &   16.174  &  \nodata  &     1.67  &     0.25   \\
     V234*  &    13:42:13.06  &     28:22:00.9  &     0.5080599  &   16.144  &    0.698  &     1.72  &     0.94   \\
     V235*  &    13:42:13.74  &     28:23:19.1  &     0.7598442  &   17.378  &    1.026  &     4.19  &     0.51   \\
      V237  &    13:42:15.71  &     28:18:16.6  &       \nodata  &   18.590  &  \nodata  &     2.01  &     0.17   \\
     V239*  &    13:42:09.87  &     28:22:14.4  &     0.5039691  &   15.671  &    0.504  &     3.14  &     0.57   \\
     V241*  &    13:42:10.76  &     28:22:37.4  &     0.5961837  &   16.044  &    0.892  &     2.87  &     1.53   \\
     V242*  &    13:42:13.15  &     28:22:24.3  &     0.5965099  &   16.027  &    0.474  &     2.41  &     1.21   \\
     V243*  &    13:42:12.32  &     28:22:14.3  &     0.6346002  &   15.735  &    0.247  &     2.16  &     0.74   \\
     V245*  &    13:42:09.83  &     28:22:58.5  &     0.2840314  &   16.231  &    0.520  &     1.87  &     1.61   \\
     V247*  &    13:42:14.74  &     28:22:15.4  &     0.6053416  &   17.277  &    0.951  &     2.28  &     1.10   \\
     V248*  &    13:42:09.64  &     28:22:48.9  &     0.5097782  &   16.501  &    0.780  &     2.86  &     1.26   \\
     V249*  &    13:42:10.41  &     28:22:47.4  &     0.6149183  &   16.006  &    0.534  &     1.95  &     1.34   \\
     V254*  &    13:42:12.40  &     28:22:53.3  &     0.6056276  &   16.192  &    0.523  &     2.12  &     1.02   \\
     V255*  &    13:42:12.65  &     28:22:43.8  &     0.5726668  &   15.709  &    0.428  &     2.13  &     1.33   \\
     V256*  &    13:42:13.02  &     28:22:57.9  &     0.3180525  &   15.891  &    0.312  &     2.21  &     0.93   \\
     V257*  &    13:42:13.03  &     28:22:29.0  &     0.6020543  &   15.858  &    0.328  &     1.45  &     1.03   \\
     V258*  &    13:42:14.29  &     28:23:30.8  &     0.7134203  &   17.522  &    1.311  &     5.22  &     0.58   \\
     V259*  &    13:42:14.55  &     28:22:54.3  &     0.3335394  &   16.720  &    0.673  &     4.15  &     0.60   \\
     V261*  &    13:42:10.10  &     28:22:39.9  &     0.4447997  &   16.122  &    0.396  &     2.77  &     0.81   \\
     V264*  &    13:42:10.86  &     28:22:29.9  &     0.3564728  &   15.969  &    0.359  &     2.10  &     1.28   \\
     V269*  &    13:42:12.76  &     28:22:31.8  &     0.3557017  &   15.278  &    0.200  &     1.60  &     1.30   \\
     V270*  &    13:42:11.94  &     28:23:31.1  &     0.6902007  &   15.672  &    0.580  &     4.09  &     0.68   \\
     V271*  &    13:42:12.16  &     28:23:18.6  &     0.6327827  &   15.516  &    0.200  &     1.85  &     0.70   \\
      V287  &    13:41:41.00  &     28:20:55.5  &       \nodata  &   20.691  &  \nodata  &     1.57  &    -0.03   \\
      V288  &    13:42:17.38  &     28:13:34.7  &       \nodata  &   18.266  &  \nodata  &     1.65  &     0.11   \\
      V289  &    13:42:12.16  &     28:19:20.9  &       \nodata  &   18.201  &  \nodata  &     1.54  &     0.18   \\
      V290  &    13:42:21.28  &     28:23:44.6  &       \nodata  &   16.735  &  \nodata  &     1.71  &     0.13   \\
      V291  &    13:42:18.05  &     28:22:38.4  &     0.0716300  &   18.052  &    0.828  &     2.95  &     0.05   \\
     V292*  &    13:42:11.17  &     28:21:53.4  &     0.2965601  &   16.567  &    0.387  &     2.93  &     0.57   \\
      V293  &    13:42:20.57  &     28:28:32.0  &       \nodata  &   18.450  &  \nodata  &     1.55  &     0.05   \\
      V294  &    13:42:13.67  &     28:25:30.0  &       \nodata  &   18.520  &  \nodata  &     1.60  &     0.30   \\
      V295  &    13:41:58.54  &     28:27:58.7  &       \nodata  &   18.793  &  \nodata  &     1.50  &     0.12   \\
      V296  &    13:41:42.98  &     28:24:03.3  &       \nodata  &   19.714  &  \nodata  &     2.48  &     0.14   \\
      V300  &    13:42:08.21  &     28:21:38.9  &   252.8500061  &   17.322  &    0.416  &     4.17  &     0.23   \\
\enddata
\end{deluxetable}

\begin{deluxetable}{lccccccc}
\tablewidth{0 pt}
\tabletypesize{\footnotesize}
\tablecaption{Parameters of Variable Stars in M~15\label{t:M15cat}}
\tablehead{
\colhead{ID} &
\colhead{$\alpha$} &
\colhead{$\delta$} &
\colhead{P} &
\colhead{$<uvm2>_{i}$} &
\colhead{$A_{uvm2}$} &
\colhead{Var} &
\colhead{SHARP} \\
\colhead{(C01)} &
\multicolumn{2}{c}{(J2000.0)} &
\colhead{(days)} &
\colhead{} &
\colhead{} &
\colhead{} &
\colhead{}}
\startdata
\hline
        V1  &    21:29:50.16  &     12:10:26.1  &     1.4377750  &   17.237  &    1.627  &     8.45  &     0.33 \\
        V2  &    21:29:46.51  &     12:10:07.2  &     0.6841473  &   17.888  &    1.238  &     4.12  &     0.32 \\
        V3  &    21:29:41.35  &     12:09:13.9  &     0.3887274  &   17.512  &    0.960  &     6.83  &     0.16 \\
        V4  &    21:29:50.62  &     12:07:17.6  &     0.3135568  &   17.310  &    0.884  &     9.23  &     0.40 \\
        V5  &    21:29:51.50  &     12:06:29.0  &     0.3842009  &   17.428  &    0.953  &     5.53  &     0.17 \\
        V6  &    21:29:59.88  &     12:11:18.9  &     0.6659238  &   18.014  &    1.741  &     3.48  &    -0.21 \\
        V7  &    21:29:58.91  &     12:11:15.7  &     0.3675779  &   17.662  &    1.009  &     4.06  &     0.04 \\
        V8  &    21:29:58.15  &     12:12:09.4  &     0.6464081  &   17.928  &    1.814  &     3.97  &     0.36 \\
        V9  &    21:29:59.19  &     12:12:21.1  &     0.7156435  &   17.712  &    1.454  &     5.11  &     0.39 \\
       V10  &    21:30:06.79  &     12:10:04.8  &     0.3863821  &   17.670  &    0.910  &     4.97  &     0.24 \\
       V11  &    21:30:09.95  &     12:09:42.1  &     0.3432007  &   17.526  &    1.034  &     5.24  &     0.26 \\
       V12  &    21:30:09.32  &     12:09:12.7  &     0.5928004  &   17.889  &    1.381  &     5.08  &     0.42 \\
       V13  &    21:30:06.89  &     12:08:54.6  &     0.5748671  &   17.885  &    1.774  &     3.60  &     0.18 \\
       V14  &    21:30:04.10  &     12:05:47.1  &     0.3819919  &   17.637  &    1.093  &     4.86  &     0.06 \\
       V15  &    21:30:03.93  &     12:04:59.2  &     0.5835864  &   17.849  &    1.786  &     5.81  &     0.09 \\
       V16  &    21:30:05.03  &     12:12:12.5  &     0.3992092  &   17.589  &    0.875  &     6.41  &     0.25 \\
       V17  &    21:30:03.88  &     12:11:52.9  &     0.4288904  &   17.566  &    0.827  &     5.79  &     0.30 \\
       V18  &    21:30:03.46  &     12:11:43.4  &     0.3677503  &   17.501  &    0.986  &     5.30  &     0.15 \\
       V19  &    21:30:05.75  &     12:12:43.4  &     0.5723480  &   17.789  &    2.454  &     5.63  &     0.07 \\
       V20  &    21:30:03.73  &     12:09:53.0  &     0.6968760  &   18.055  &    1.709  &     3.08  &     0.36 \\
      V21*  &    21:30:00.57  &     12:09:05.1  &     0.6489394  &   17.674  &    1.404  &     4.68  &     0.55 \\
       V22  &    21:29:35.74  &     12:09:14.0  &     0.7202495  &   17.805  &    1.658  &     4.87  &     0.09 \\
       V23  &    21:30:11.17  &     12:14:19.6  &     0.6326446  &   17.821  &    1.418  &     5.13  &     0.40 \\
       V24  &    21:29:50.97  &     12:09:55.5  &     0.3697256  &   17.596  &    0.971  &     6.15  &     0.20 \\
       V25  &    21:30:18.84  &     12:09:53.8  &     0.6653631  &   18.021  &    1.484  &     3.52  &     0.30 \\
       V26  &    21:29:59.66  &     12:15:33.9  &     0.4023091  &   17.742  &    0.731  &     5.90  &     0.09 \\
       V27  &    21:30:13.28  &     12:14:12.3  &       \nodata  &   20.170  &  \nodata  &     1.33  &     0.30 \\
       V29  &    21:30:09.21  &     12:13:34.6  &     0.5747560  &   18.091  &    1.419  &     5.22  &     0.14 \\
      V30*  &    21:29:47.02  &     12:09:58.6  &     0.4060355  &   16.471  &    0.222  &     1.85  &     0.51 \\
       V31  &    21:29:50.46  &     12:14:06.5  &     0.4081596  &   17.670  &    0.808  &     5.26  &     0.18 \\
       V32  &    21:29:54.74  &     12:11:49.5  &     0.6053026  &   17.554  &    1.086  &     5.36  &     0.21 \\
      V33*  &    21:29:55.47  &     12:09:33.7  &     0.5839413  &   17.300  &    1.339  &     3.97  &     0.61 \\
       V34  &    21:29:54.47  &     12:09:07.2  &     2.0373480  &   17.651  &    0.688  &     4.65  &     0.28 \\
       V35  &    21:29:55.99  &     12:07:18.7  &     0.3839766  &   17.710  &    0.997  &     6.60  &     0.03 \\
       V36  &    21:29:56.36  &     12:08:40.8  &     0.6241769  &   17.897  &    1.695  &     4.01  &     0.46 \\
       V37  &    21:29:56.53  &     12:08:45.1  &     0.2874888  &   17.193  &    0.745  &     5.20  &     0.42 \\
       V38  &    21:29:58.80  &     12:07:36.3  &     0.3752608  &   17.532  &    1.033  &     4.79  &     0.27 \\
      V39*  &    21:29:59.66  &     12:07:58.0  &     0.3895853  &   17.553  &    0.628  &     4.20  &     0.61 \\
       V40  &    21:30:07.22  &     12:08:06.4  &     0.3773258  &   17.589  &    1.058  &     5.25  &     0.11 \\
       V41  &    21:30:02.55  &     12:09:07.6  &     0.3917472  &   17.427  &    0.735  &     4.34  &     0.44 \\
       V42  &    21:30:13.71  &     12:09:27.1  &     0.3601973  &   17.637  &    1.003  &     6.42  &     0.15 \\
       V43  &    21:30:26.57  &     12:11:48.0  &     0.3959861  &   17.649  &    0.923  &     4.69  &     0.19 \\
       V44  &    21:30:04.43  &     12:10:06.7  &     0.5959346  &   17.842  &    1.463  &     5.15  &     0.21 \\
      V45*  &    21:30:02.76  &     12:09:31.4  &     0.6775171  &   17.692  &    0.557  &     3.70  &     0.75 \\
      V47*  &    21:30:01.24  &     12:09:59.1  &     0.6874813  &   16.089  &    0.454  &     2.73  &     0.73 \\
       V48  &    21:30:02.22  &     12:12:32.8  &     0.3649691  &   17.517  &    1.117  &     5.76  &     0.26 \\
       V49  &    21:30:00.89  &     12:12:49.1  &     0.6552283  &   16.647  &    0.584  &     3.30  &     0.27 \\
       V50  &    21:30:09.37  &     12:11:43.2  &     0.2980822  &   17.467  &    1.017  &     7.26  &     0.22 \\
       V52  &    21:30:11.34  &     12:09:41.7  &     0.5756440  &   18.097  &    1.400  &     4.20  &     0.28 \\
       V53  &    21:29:51.99  &     12:08:11.3  &     0.4141384  &   17.528  &    0.827  &     4.85  &     0.18 \\
      V54*  &    21:29:58.93  &     12:11:30.3  &     0.3995029  &   17.475  &    0.652  &     3.43  &     0.60 \\
      V55*  &    21:30:02.70  &     12:09:43.9  &     0.7485716  &   18.110  &    1.138  &     3.61  &     0.59 \\
       V56  &    21:30:02.15  &     12:10:03.7  &     0.5703183  &   16.188  &    0.504  &     2.77  &     0.40 \\
       V57  &    21:30:03.35  &     12:09:07.9  &     0.3492579  &   17.480  &    1.143  &     4.90  &     0.08 \\
      V58*  &    21:29:54.46  &     12:10:10.6  &     0.4071922  &   17.519  &    0.649  &     3.07  &     0.52 \\
       V61  &    21:29:53.67  &     12:09:20.6  &     0.3996260  &   17.635  &    0.919  &     4.24  &     0.22 \\
      V62*  &    21:29:53.35  &     12:10:40.8  &     0.3772887  &   17.388  &    0.873  &     4.72  &     0.66 \\
      V64*  &    21:29:55.08  &     12:10:21.9  &     0.3642009  &   17.338  &    0.864  &     3.75  &     0.78 \\
       V65  &    21:29:51.28  &     12:09:23.3  &     0.7181118  &   17.870  &    1.198  &     4.01  &     0.48 \\
       V66  &    21:29:53.61  &     12:08:10.1  &     0.3793565  &   17.553  &    0.788  &     4.86  &     0.38 \\
       V67  &    21:29:52.35  &     12:09:51.8  &     0.3674503  &   17.824  &    1.069  &     5.53  &    -0.25 \\
       V69  &    21:29:55.79  &     12:09:35.9  &     0.5869104  &   18.103  &    1.266  &     2.32  &    -0.19 \\
      V70*  &    21:29:55.92  &     12:09:42.9  &     0.3676001  &   17.083  &    0.580  &     2.91  &     1.09 \\
      V71*  &    21:29:55.92  &     12:09:50.3  &     0.3741176  &   16.891  &    0.457  &     1.60  &     2.08 \\
       V74  &    21:30:00.84  &     12:08:37.0  &     0.2960204  &   17.430  &    1.159  &     5.28  &     0.23 \\
       V78  &    21:29:57.69  &     12:10:50.0  &     0.6646888  &   16.111  &    0.411  &     2.94  &     0.41 \\
       V80  &    21:29:55.00  &     12:09:36.2  &     0.6640881  &   16.318  &    0.404  &     2.44  &     0.43 \\
      V88*  &    21:29:58.38  &     12:10:29.2  &     0.6835572  &   16.350  &    0.758  &     2.02  &     1.87 \\
       V96  &    21:30:09.40  &     12:13:38.1  &     0.3963543  &   17.874  &    0.706  &     4.93  &     0.32 \\
       V97  &    21:29:52.82  &     12:10:30.9  &     0.6963688  &   17.953  &    1.210  &     3.77  &     0.28 \\
       V99  &    21:30:00.30  &     12:13:15.1  &     0.2908204  &   17.330  &    0.370  &     4.09  &     0.15 \\
     V100*  &    21:29:59.26  &     12:09:23.7  &     0.4063688  &   16.962  &    0.451  &     2.29  &     1.25 \\
     V102*  &    21:30:03.04  &     12:10:31.6  &     0.7595775  &   17.751  &    0.663  &     3.08  &     0.74 \\
      V103  &    21:29:41.24  &     12:05:27.0  &     0.3682380  &   17.542  &    1.112  &     5.56  &     0.08 \\
      V113  &    21:29:58.73  &     12:05:52.9  &     0.4059446  &   17.430  &    0.353  &     3.22  &     0.07 \\
      V114  &    21:29:58.13  &     12:10:52.2  &       \nodata  &   16.867  &  \nodata  &     1.72  &     0.20 \\
     V116*  &    21:29:59.34  &     12:09:12.8  &     0.6137645  &   17.356  &    1.131  &     3.43  &     0.90 \\
      V118  &    21:29:59.53  &     12:10:55.4  &     0.2999837  &   16.565  &    0.352  &     2.94  &     0.22 \\
      V122  &    21:30:15.81  &     12:10:26.5  &       \nodata  &   16.680  &  \nodata  &     1.45  &     0.06 \\
      V123  &    21:29:42.95  &     12:09:53.3  &       \nodata  &   19.376  &  \nodata  &     1.87  &     0.29 \\
      V124  &    21:29:50.12  &     12:06:40.4  &       \nodata  &   19.599  &  \nodata  &     1.72  &     0.42 \\
      V126  &    21:30:10.42  &     12:10:05.4  &       \nodata  &   20.523  &  \nodata  &     0.99  &    -0.26 \\
      V127  &    21:30:21.06  &     12:11:32.7  &       \nodata  &   20.951  &  \nodata  &     1.36  &    -0.05 \\
      V156  &    21:29:39.39  &     12:11:43.2  &       \nodata  &   19.255  &  \nodata  &     1.58  &     0.01 \\
     V170*  &    21:29:58.77  &     12:09:21.6  &     0.6779041  &   17.226  &    0.834  &     2.79  &     1.43 \\
      V182  &    21:30:02.87  &     12:10:08.8  &     0.3913914  &   17.928  &    1.025  &     4.25  &    -0.39 \\
\enddata
\end{deluxetable}

\begin{deluxetable}{lcccc}
\tablewidth{0 pt}
\tabletypesize{\footnotesize}
\tablecaption{Photometry of Variable Stars in M~3\label{t:M3phot}}
\tablehead{
\colhead{ID} &
\colhead{JD} &
\colhead{$uvm2$} &
\colhead{$\sigma_{uvm2}$}}
\startdata
\hline
        V1  &  2455998.8316  &   16.829  &    0.063 \\
        V1  &  2455999.0348  &   18.263  &    0.080 \\
        V1  &  2455999.0990  &   18.147  &    0.108 \\
        V1  &  2455999.2981  &   15.987  &    0.054 \\
        V1  &  2456002.9276  &   15.782  &    0.046 \\
        V1  &  2456006.7371  &   17.512  &    0.080 \\
        V1  &  2456010.7942  &   16.592  &    0.061 \\
        V1  &  2456010.9301  &   17.498  &    0.032 \\
        V1  &  2456019.3608  &   17.965  &    0.100 \\
        V1  &  2456022.7672  &   16.673  &    0.048 \\
\enddata
\tablenotetext{1}{Table \ref{t:M3phot} is presented in its entirety in the electronic edition of the Astronomical Journal. A portion is how here for guidance regarding
its form and content.}
\end{deluxetable}

\begin{deluxetable}{lcccc}
\tablewidth{0 pt}
\tabletypesize{\footnotesize}
\tablecaption{Photometry of Variable Stars in M~15\label{t:M15phot}}
\tablehead{
\colhead{ID} &
\colhead{JD} &
\colhead{$uvm2$} &
\colhead{$\sigma_{uvm2}$}}
\startdata
\hline
        V1  &  2455702.9226  &   16.927  &    0.069 \\
        V1  &  2456403.3904  &   17.677  &    0.071 \\
        V1  &  2456410.6693  &   17.604  &    0.052 \\
        V1  &  2456419.4170  &   17.874  &    0.059 \\
        V1  &  2456426.6931  &   18.104  &    0.076 \\
        V1  &  2456434.7068  &   16.605  &    0.042 \\
        V1  &  2456439.0323  &   16.641  &    0.048 \\
        V1  &  2456450.9921  &   17.702  &    0.092 \\
        V1  &  2456459.2694  &   17.106  &    0.056 \\
        V1  &  2456467.0026  &   17.948  &    0.072 \\
 \enddata
\tablenotetext{1}{Table \ref{t:M15phot} is presented in its entirety in the electronic edition of the Astronomical Journal. A portion is how here for guidance regarding
its form and content.}
\end{deluxetable}

\begin{figure}[h]
\begin{center}
\includegraphics[scale=1]{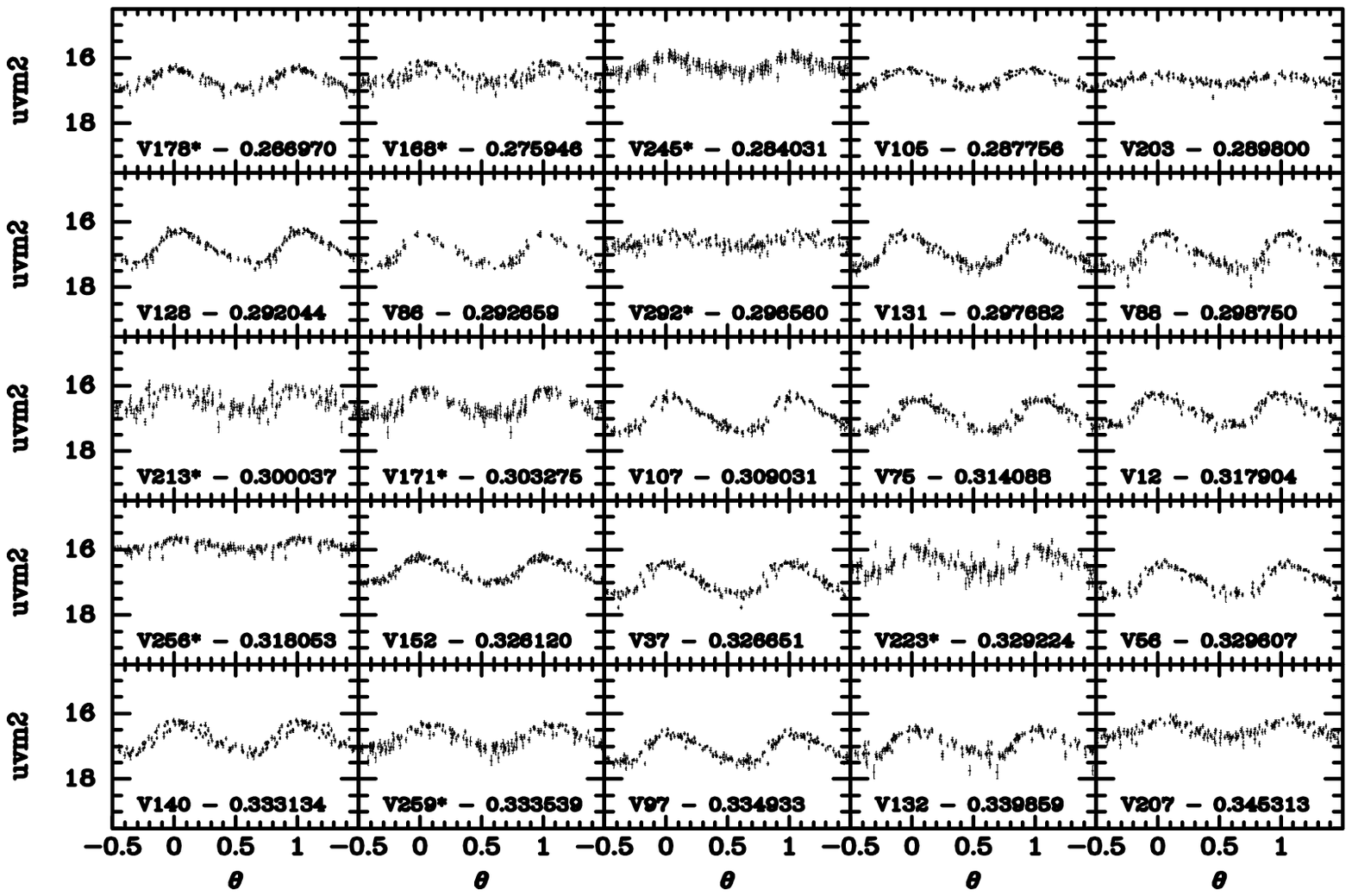}
\end{center}
\caption{$uvm2$ light curves for the 199 RR Lyrae stars identified from the Swift data of the globular cluster M~3.  Stars marked with asterisks represent
those with SHARP values greater than 0.5.  These are stars that are mostly in the core of M~3, identified as potential variable stars from the C01 catalog.  They
are likely blends, with concomitant elevated magnitudes and dampened pulsational amplitudes.  They have been excluded from
our detailed analysis.\label{f:M3RRL0}}
\end{figure}

\begin{figure}[h]

\begin{center}
\includegraphics[scale=1]{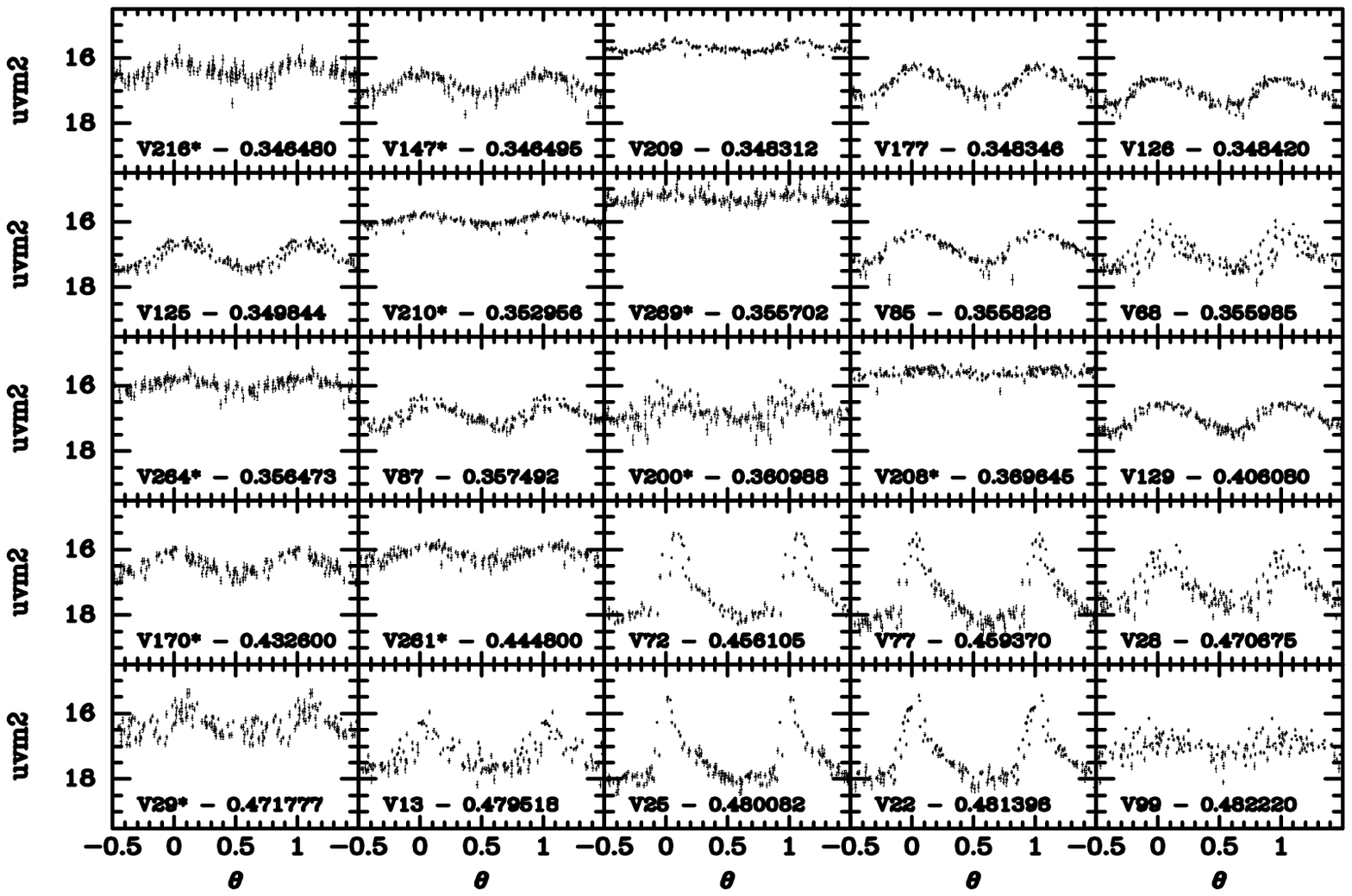}
\end{center}
\caption{$uvm2$ light curves continued.\label{f:M3RRL1}}
\end{figure}

\begin{figure}[h]

\begin{center}
\includegraphics[scale=1]{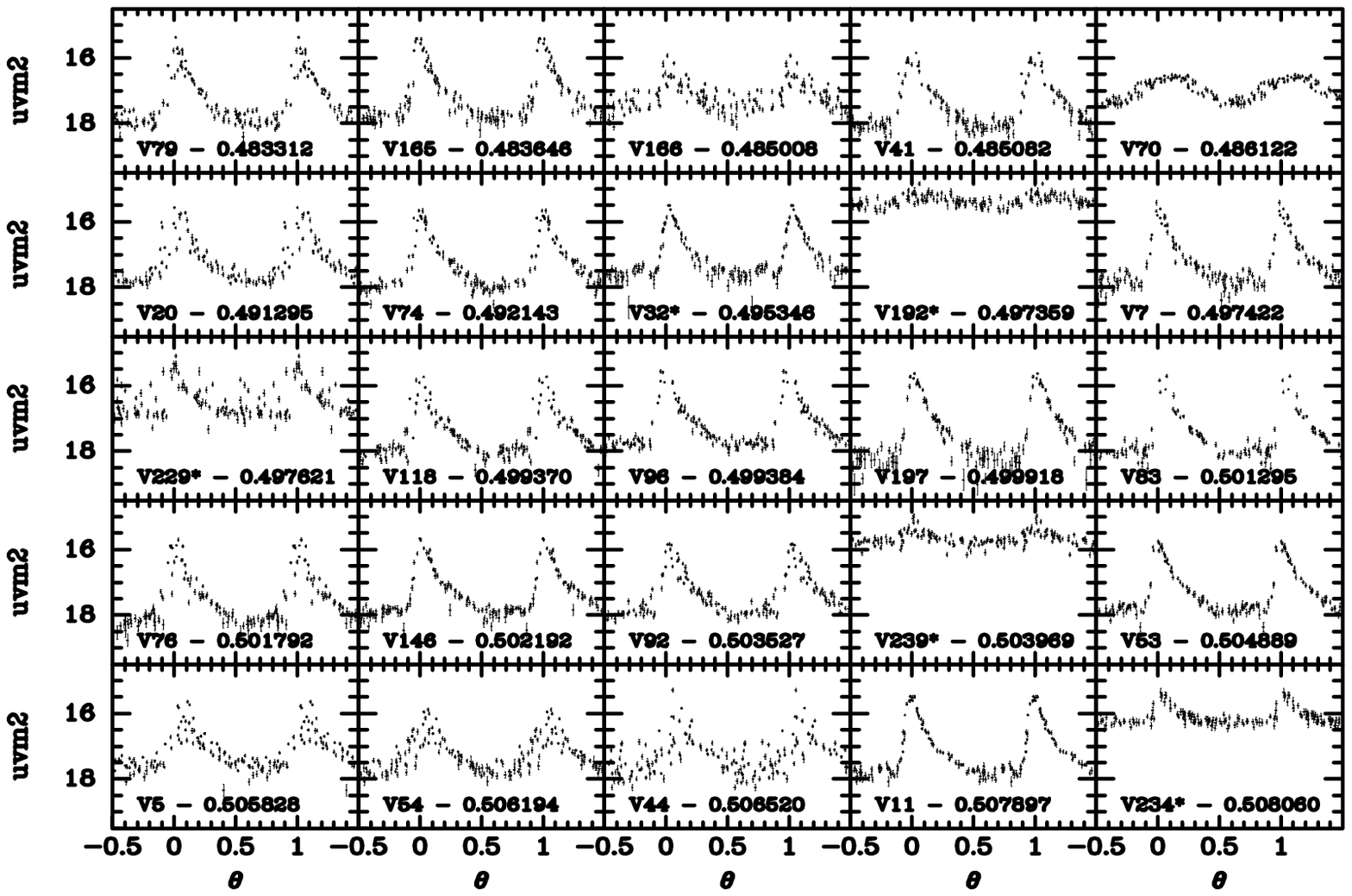}
\end{center}
\caption{$uvm2$ light curves continued.\label{f:M3RRL2}}
\end{figure}

\begin{figure}[h]

\begin{center}
\includegraphics[scale=1]{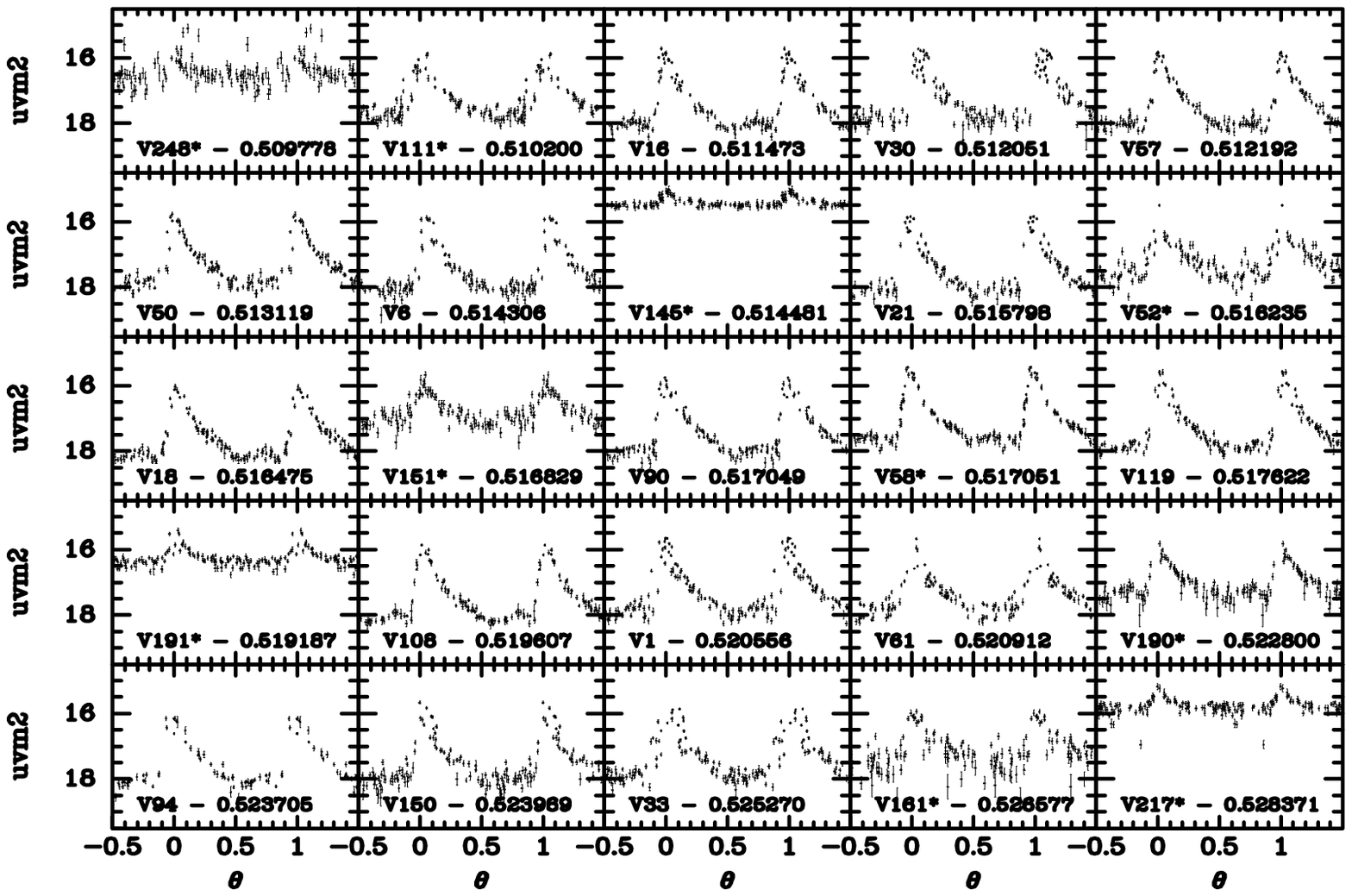}
\end{center}
\caption{$uvm2$ light curves continued.\label{f:M3RRL3}}
\end{figure}

\begin{figure}[h]

\begin{center}
\includegraphics[scale=1]{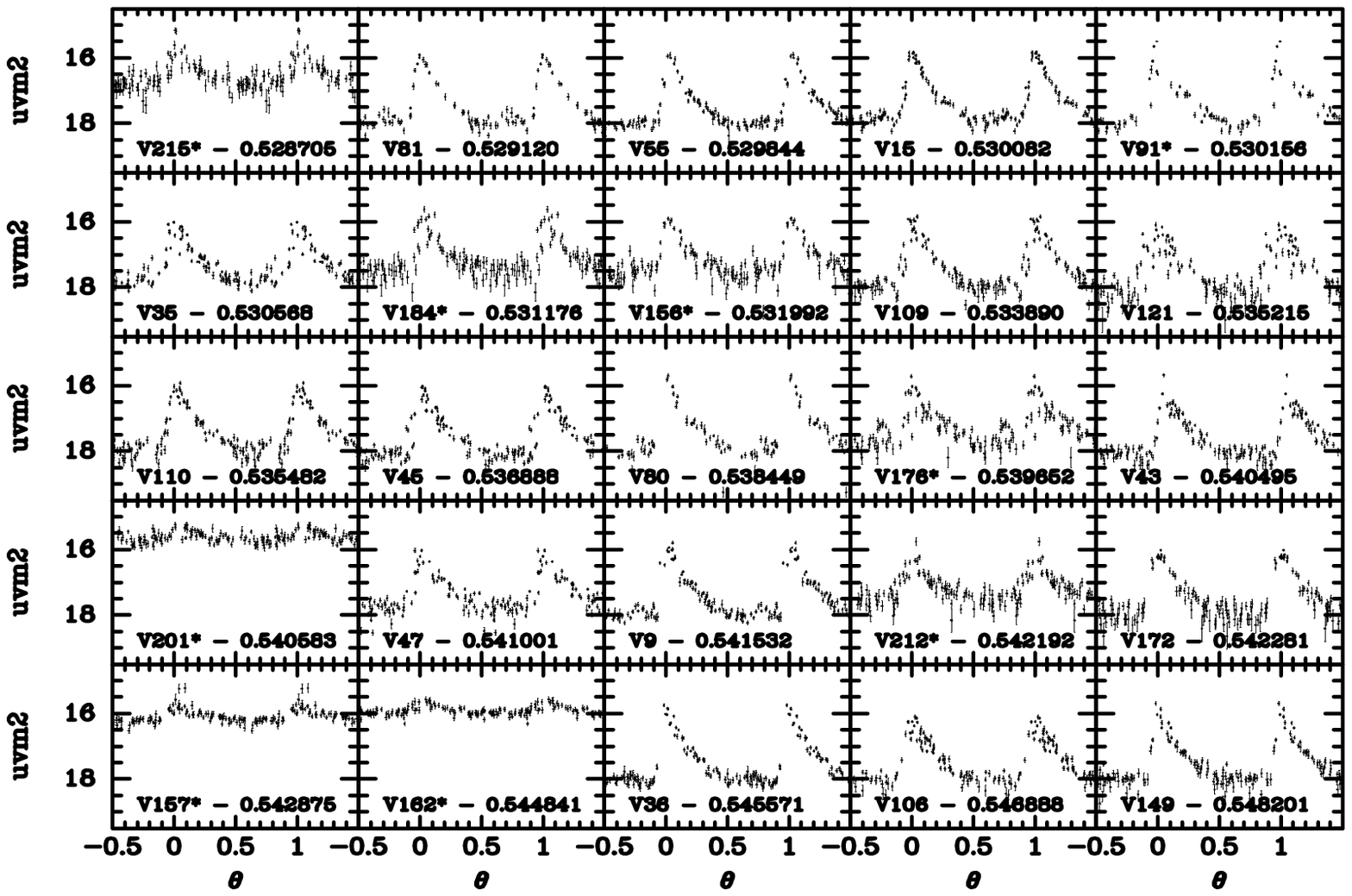}
\end{center}
\caption{$uvm2$ light curves continued.\label{f:M3RRL4}}
\end{figure}

\begin{figure}[h]

\begin{center}
\includegraphics[scale=1]{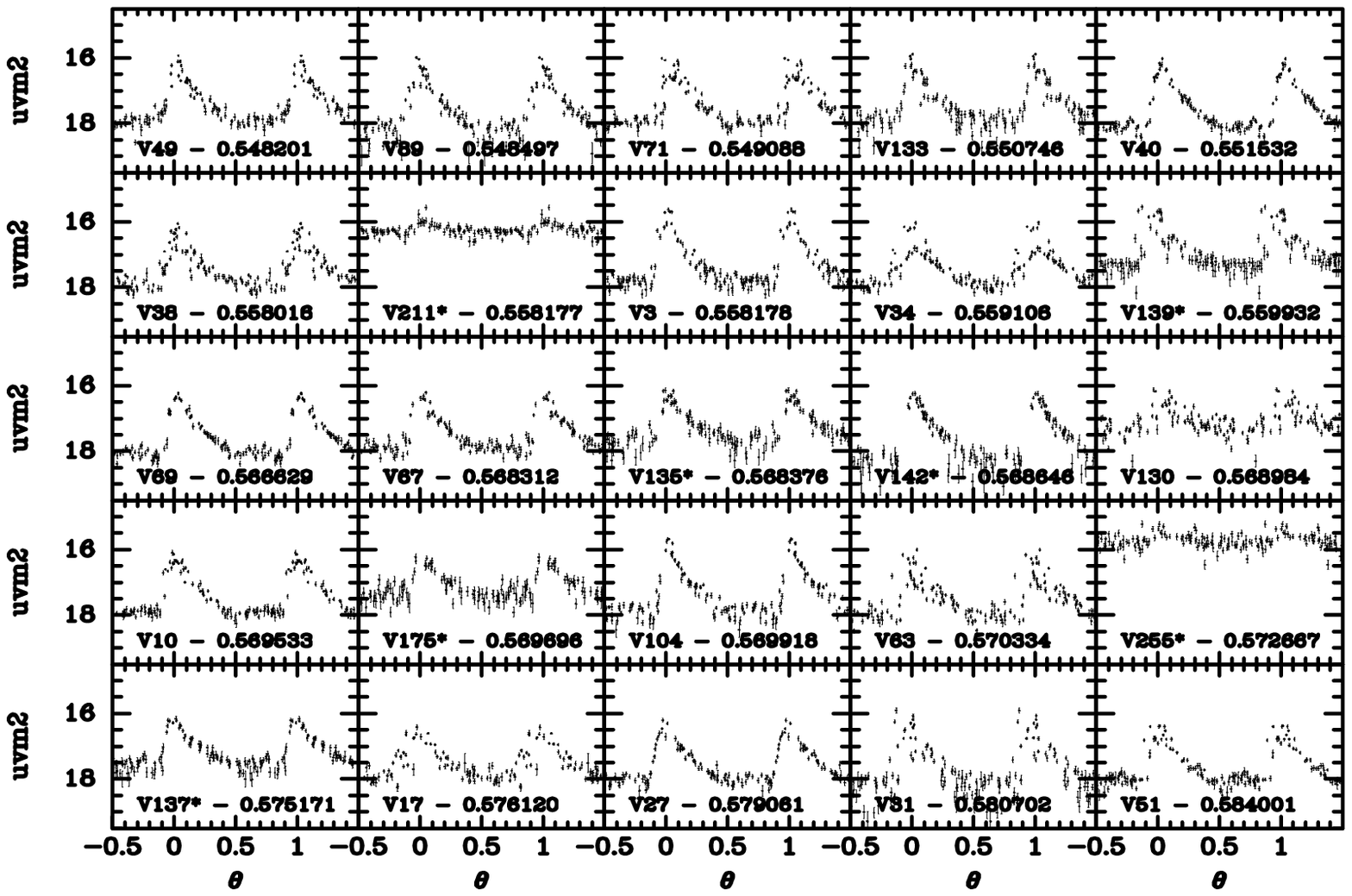}
\end{center}
\caption{$uvm2$ light curves continued.\label{f:M3RRL5}}
\end{figure}

\begin{figure}[h]

\begin{center}
\includegraphics[scale=1]{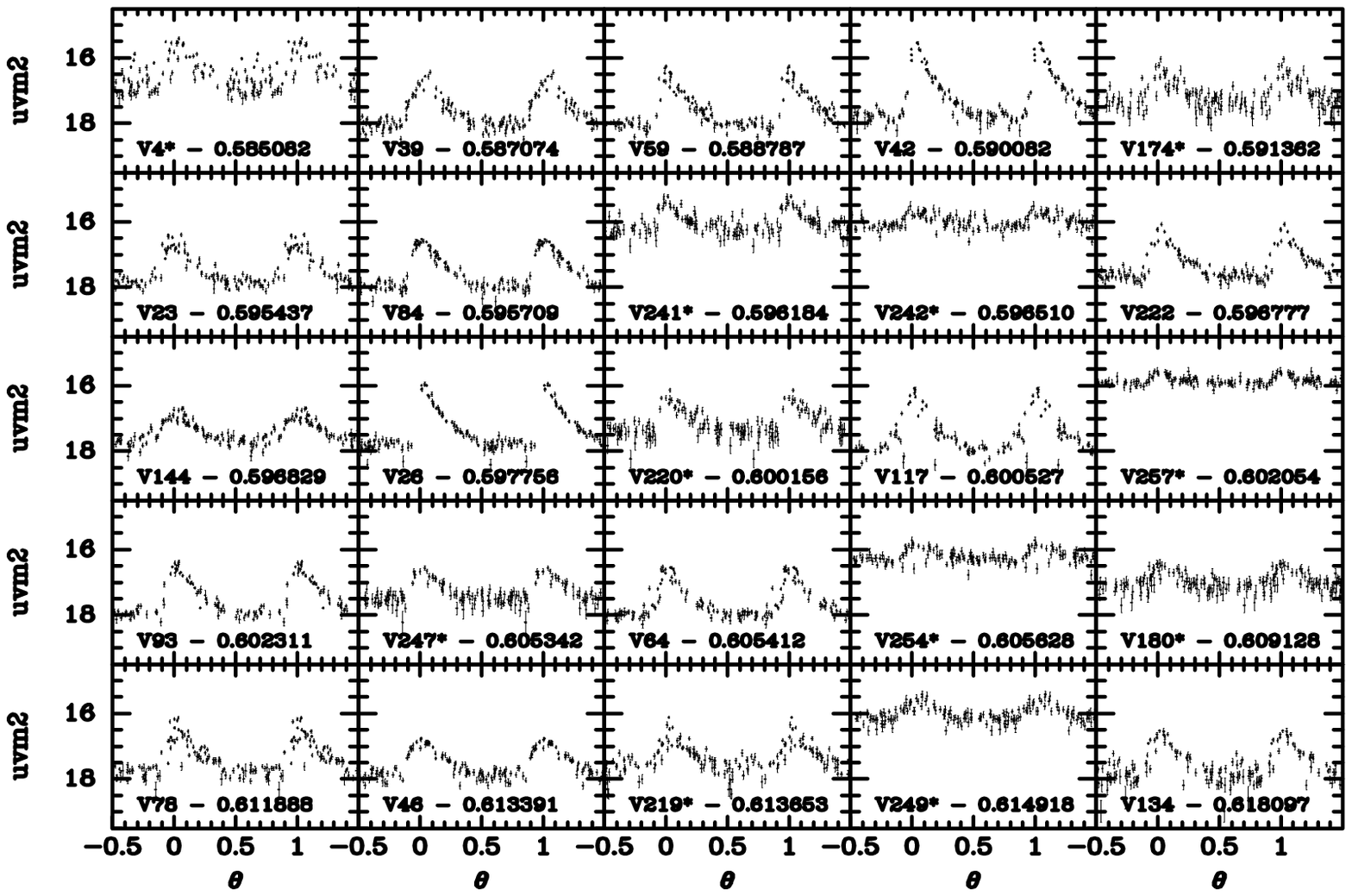}
\end{center}
\caption{$uvm2$ light curves continued.\label{f:M3RRL6}}
\end{figure}

\begin{figure}[h]

\begin{center}
\includegraphics[scale=1]{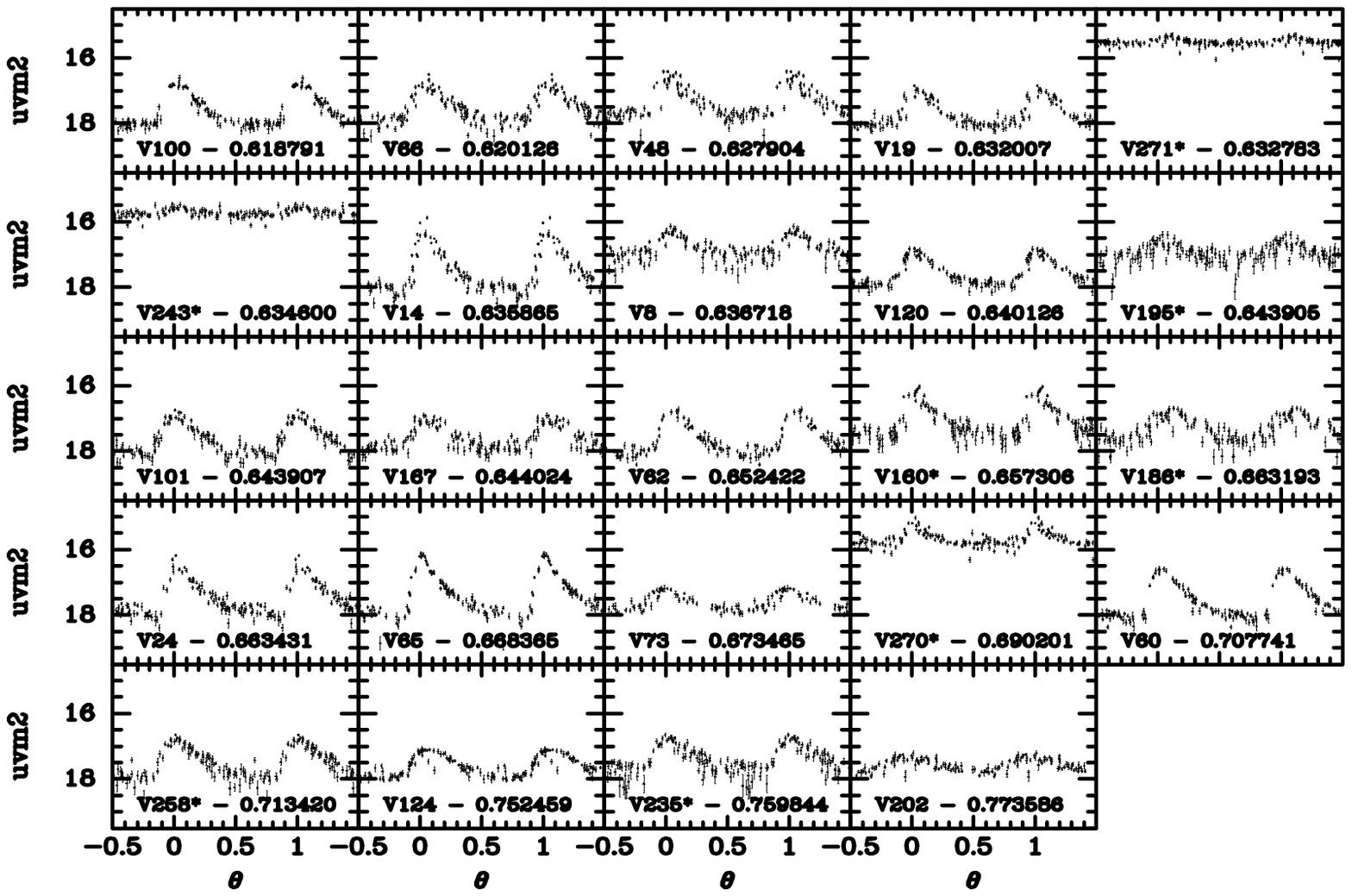}
\end{center}
\caption{$uvm2$ light curves continued.\label{f:M3RRL7}}
\end{figure}

\begin{figure}[h]
\begin{center}
\includegraphics[scale=1]{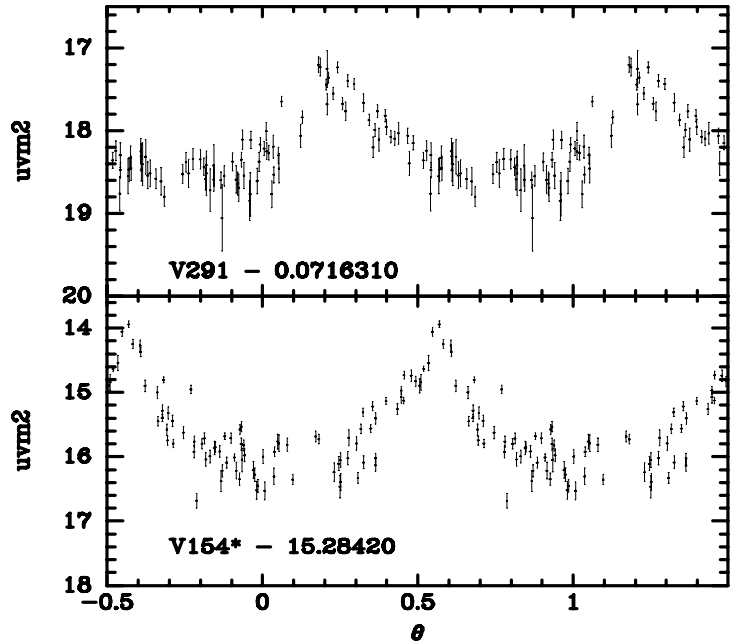}
\end{center}
\caption{$uvm2$ light curves for the SX Phoenicis star V291 (top) and the classical Cepheid V154 (bottom) in the globular cluster M~3.\label{f:M3other}}
\end{figure}

\begin{figure}[h]
\begin{center}
\includegraphics[scale=1]{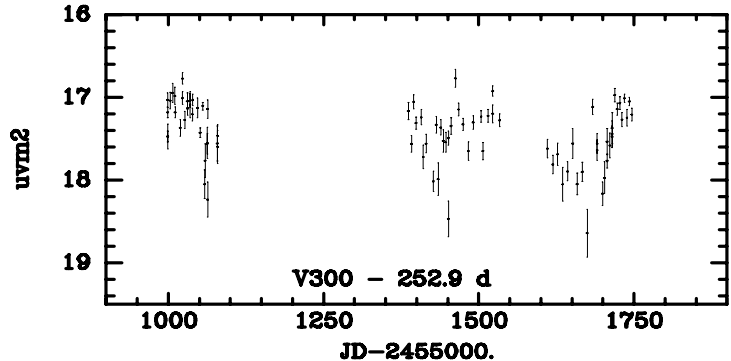}
\end{center}
\caption{$uvm2$ light curve for a possible long-term or irregular variable discovered in M~3.\label{f:M3long}}
\end{figure}

\begin{figure}[h]
\begin{center}
\includegraphics[scale=1]{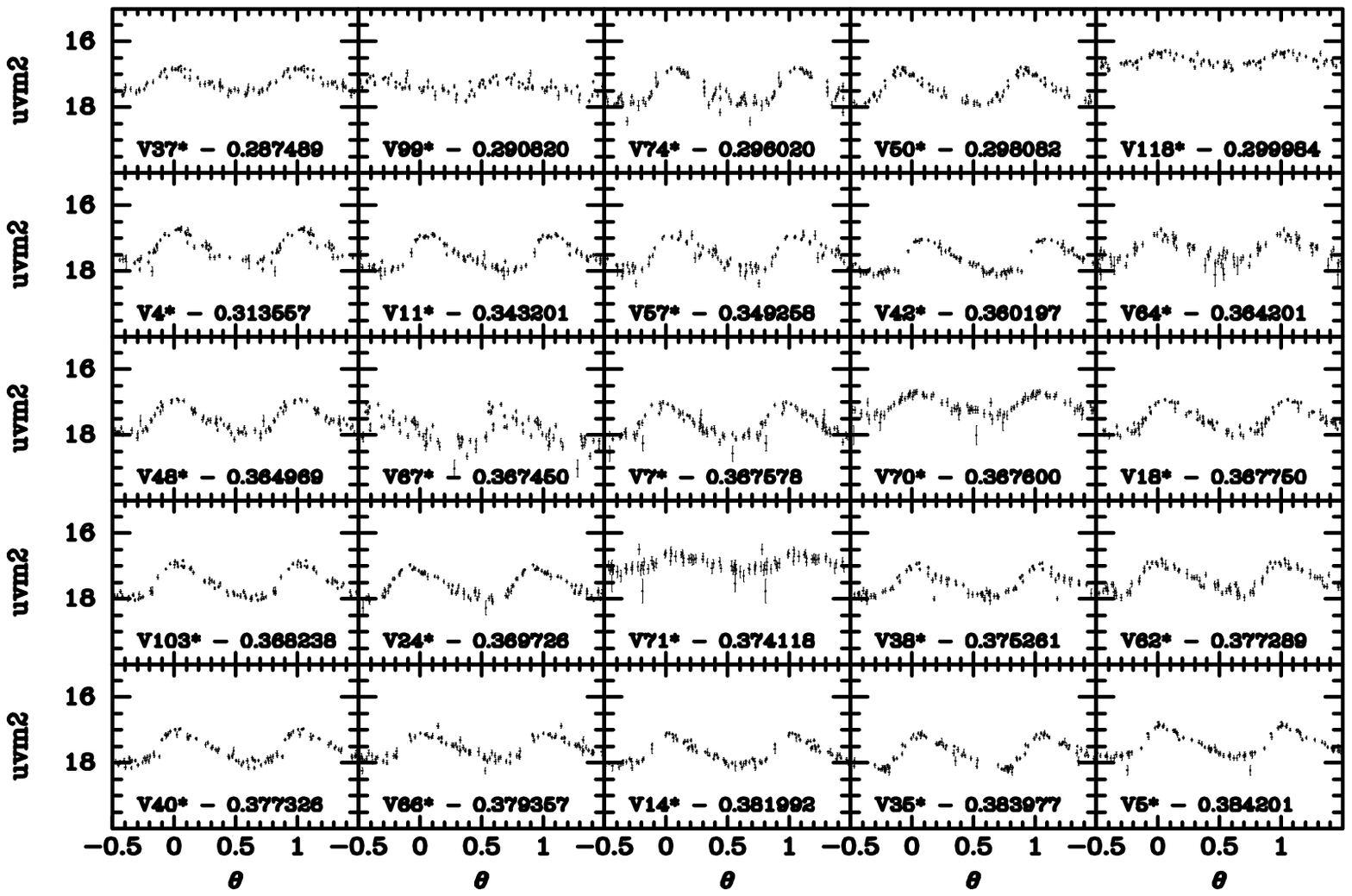}
\end{center}
\caption{$uvm2$ light curves for the 76 RR Lyrae stars identified from the Swift data of the globular cluster M~15.  Stars marked with asterisks represent
those with SHARP values greater than 0.5.  These are stars that are mostly in the core of M~15, identified as potential variable stars from the C01 catalog.  They
are likely blends, with concomittant elevated magnitudes and dampened pulsational amplitudes.  They have been excluded from
our detailed analysis.\label{f:M15RRL0}}
\end{figure}

\begin{figure}[h]

\begin{center}
\includegraphics[scale=1]{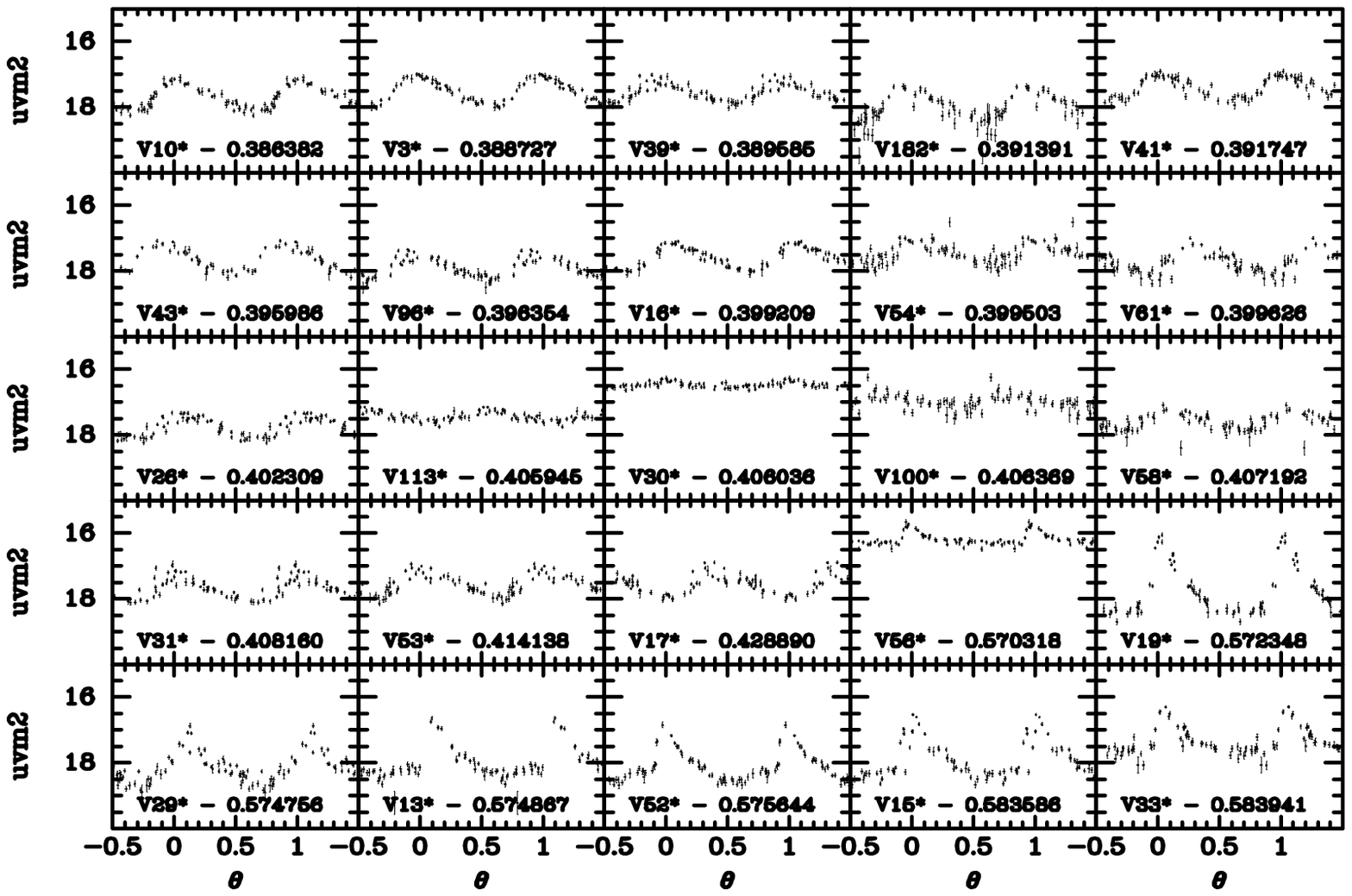}
\end{center}
\caption{$uvm2$ light curves continued.\label{f:M15RRL1}}
\end{figure}

\begin{figure}[h]

\begin{center}
\includegraphics[scale=1]{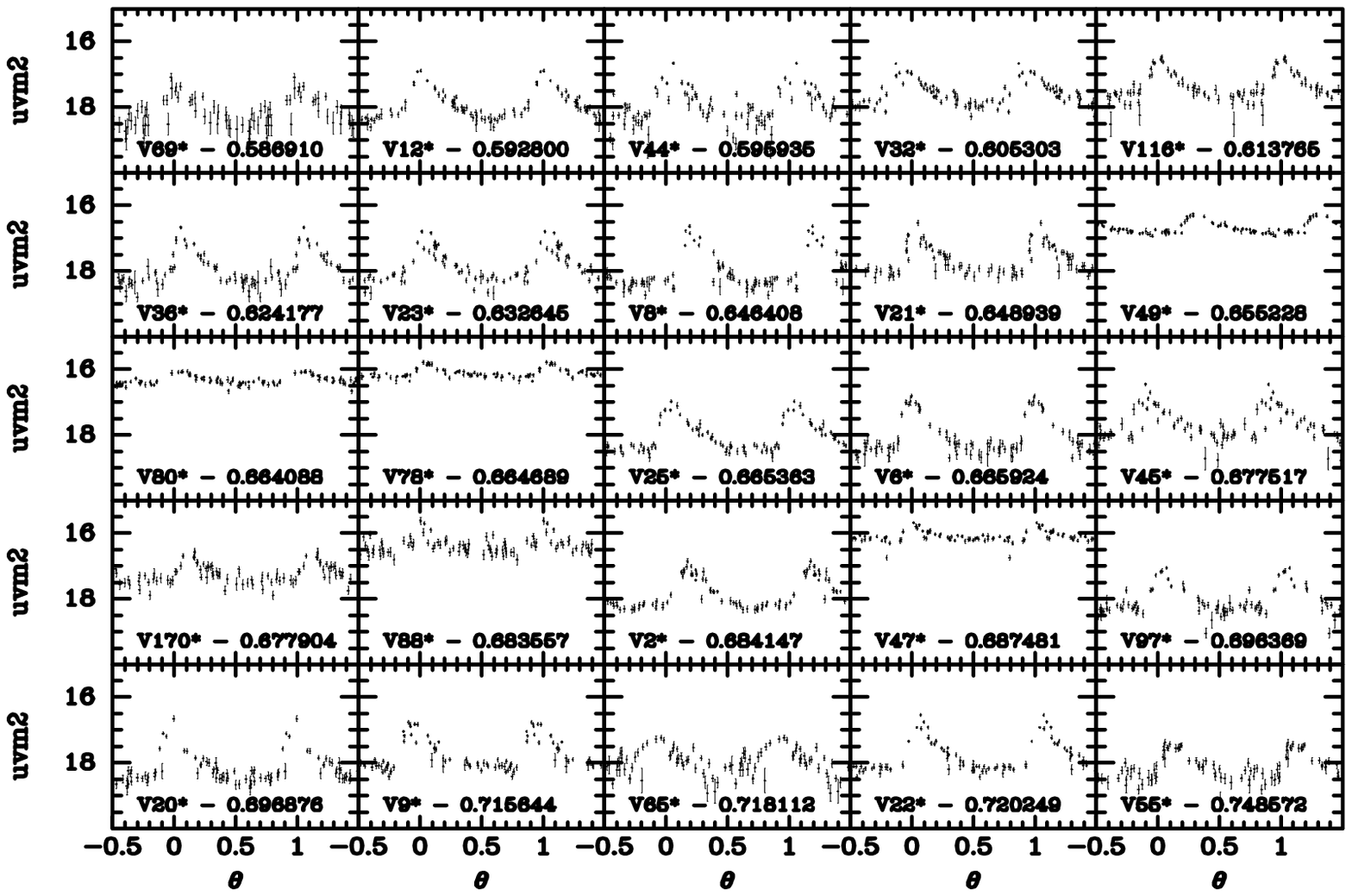}
\end{center}
\caption{$uvm2$ light curves continued.\label{f:M15RRL2}}
\end{figure}

\begin{figure}[h]

\begin{center}
\includegraphics[scale=1]{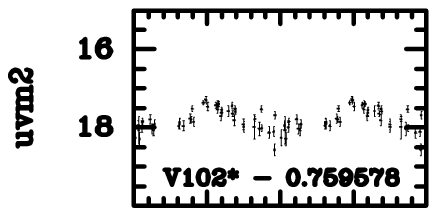}
\end{center}
\caption{$uvm2$ light curves continued.\label{f:M15RRL3}}
\end{figure}

\begin{figure}[h]
\begin{center}
\includegraphics[scale=1]{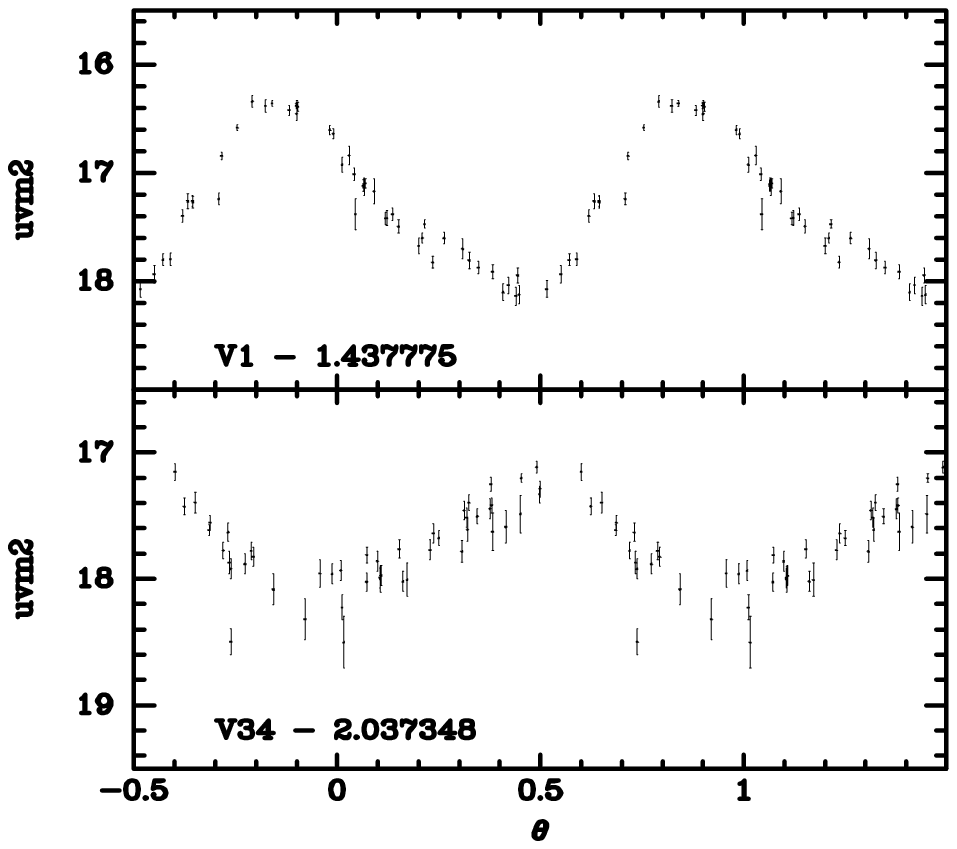}
\end{center}
\caption{$uvm2$ light curves for two anomalous Cepheids in the globular cluster M~15.\label{f:M15other}}
\end{figure}

\subsection{Comparison to Previous Investigations}
\label{ss:comp}

Of the 290 variable stars and variable star candidates listed in C01 for M~3, we cross-identify 217 to our UVOT data.  Of these, we were able to fit periods to 201.
Figure \ref{f:M3periodcomp} compares the periods derived in this study of M~3 against those compiled in C01.  As can be seen, the periods
line up almost exactly, with 174 of the 201 comparisons being within .0001 days and 190 being within .001 days.  Of the stars that have substantial differences
in periods:

\begin{enumerate}

\item V79 and V99 have periods of 0.3581 and 0.3609 days, respectively, in C01.  Our analysis finds periods of 0.4833 and 0.48222 days, respectively.  Both
are listed as double-mode pulsators, however and it is likely that we have measured the fundamental mode. This would be consistent with a mode ratio
of 0.74-0.75 (Popielski et al. 2000).

\item V162 is listed in C01 as a non-variable star that is part of a triple system.  It does show periodicity in our data (although the variability is 
low and the SHARP value high).  We list it among the RR Lyrae stars but note that it is probably a blended multiple star system with at least one
component having RR Lyrae-like pulsations.  Because of its high SHARP value, it is excluded from our analysis.

\item V217 is listed in C01 as an RR Lyrae variable with no period given.  Although the object has moderate variability and a high SHARP value, we do find
a periodicity of 0.528 days with an RRab shaped light curve.

\end{enumerate}

Of the 16 stars we match to C01 but are unable to fit periods to, none have a variability index greater than 3.0. Of these stars:

\begin{enumerate}

\item Six stars listed in C01 as nonvariable are confirmed as non-variable by our data.

\item Six stars listed in C01 as SX Phoenicis stars are measured as non-variable in our data.  SX Phoenicis stars tend to be faint with rapid pulsations
and small amplitudes.  Our data are not optimized for their detection so it is not surprising that we can not confirm their variation.  The only SX Phoenicis
star we do confirm -- V291 -- has a large amplitude and was identified by {\it a priori} knowledge of the period.

\item V138, listed by C01 as a semi-regular variable, shows little variation in our data, with a dispersion of only 1.1 times the formal scatter.  This is not surprising
as these stars are very faint in the UV.

\item V287, which C01 are unable to classify, shows little variation in our data, with a dispersion of only 1.6 times the formal scatter

\item V290, listed in C01 as a double-mode RR Lyrae, shows no clear minimum in the phase dispersion diagram. It is possible our analysis is being confounded by the
double mode.

\item V296, listed by C01 as an eclipsing binary with a period of .446 days, shows no clear minimum in the phase dispersion diagram. It does have a weak minimum
on .377 day cycle.

\end{enumerate}

We identify one potential new variable star in M~3 -- marked as V300 in Table \ref{t:M3cat}.  This star shows significant variation over the course of the observing campaign
with a very crude period of 253 days (see Figure \ref{f:M3long}).  We note, however, that our data span just over one pulsation cycle and show multiple minima in the 200-250 day range as well
as in other ranges of periodicity.  If this is indeed
a real variable, it could be pulsating on a much more irregular or higher-order cycle.  Further investigation is warranted.

\begin{figure}[h]
\begin{center}
\includegraphics[scale=1]{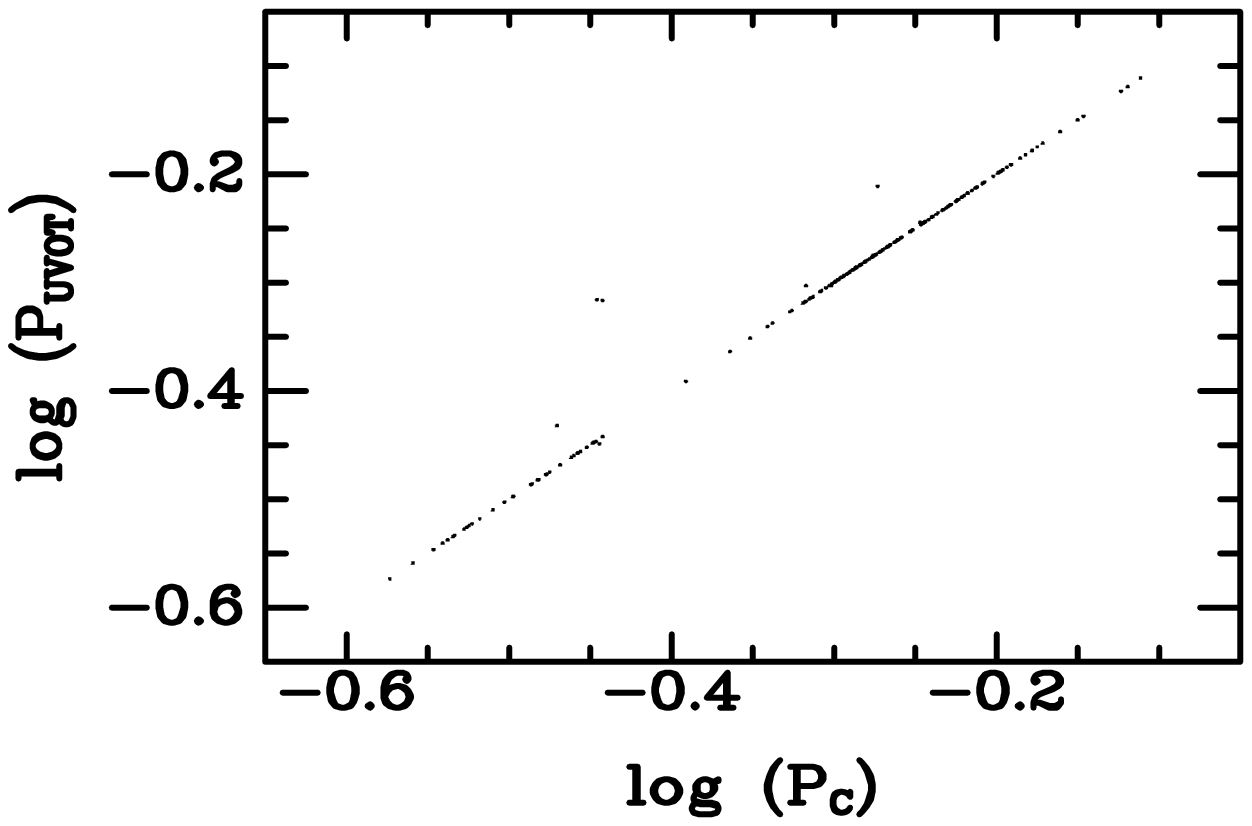}
\end{center}
\caption{A comparison of the periods derived in this study (ordinate) against the compilation of C01 (abscissa) for the globular cluster M~3.  The outliers
are discussed in the text.\label{f:M3periodcomp}}
\end{figure}

Of the 182 variable stars and variable star candidates identified by C01 in M~15, we are able to identify counterparts to 85.  Of these, we were able to fit periods to 77.
Figure \ref{f:M15periodcomp} compares the periods derived in this study of M~15 against those compiled in C01.  As can be seen, the periods
line up almost exactly, with 47 of the 77 comparisons being within .0001 days and 68 being within .001 days.  The only star which has a significant difference
in period is V34.  This star is listed by C08 as a possible eclipsing binary with a period of 1.1591 days and shows a somewhat irregular light curve.  We find a period of 2.037 days
and show a sawtooth pattern to the light curve, which indicates that this could be an anomalous Cepheid.  We also identify one new RRc star, which we label V182.

Of the eight stars we match to C01 but are unable to fit periods to, none have a variability index greater than 3.0,  Of these stars:

\begin{enumerate}

\item Five stars listed in C01 as of uncertain nature (V122, V123, V124, V126 and V127) show up as non-variable in our data.  We note that these stars are faint
and have small amplitudes.

\item V27 is listed as non-variable and we confirm this, measure a dispersion of 1.33 times the formal error.

\item We do not detect any variability in the SX Phoenicis star V156 to a limit of 1.56 times the formal error. This is not surprising given its faintness and small amplitude.

\item The RR Lyrae variable V114 has C01 position that lies between two point source in our data.  Neither shows significant variability.

\end{enumerate}

\begin{figure}[h]
\begin{center}
\includegraphics[scale=1]{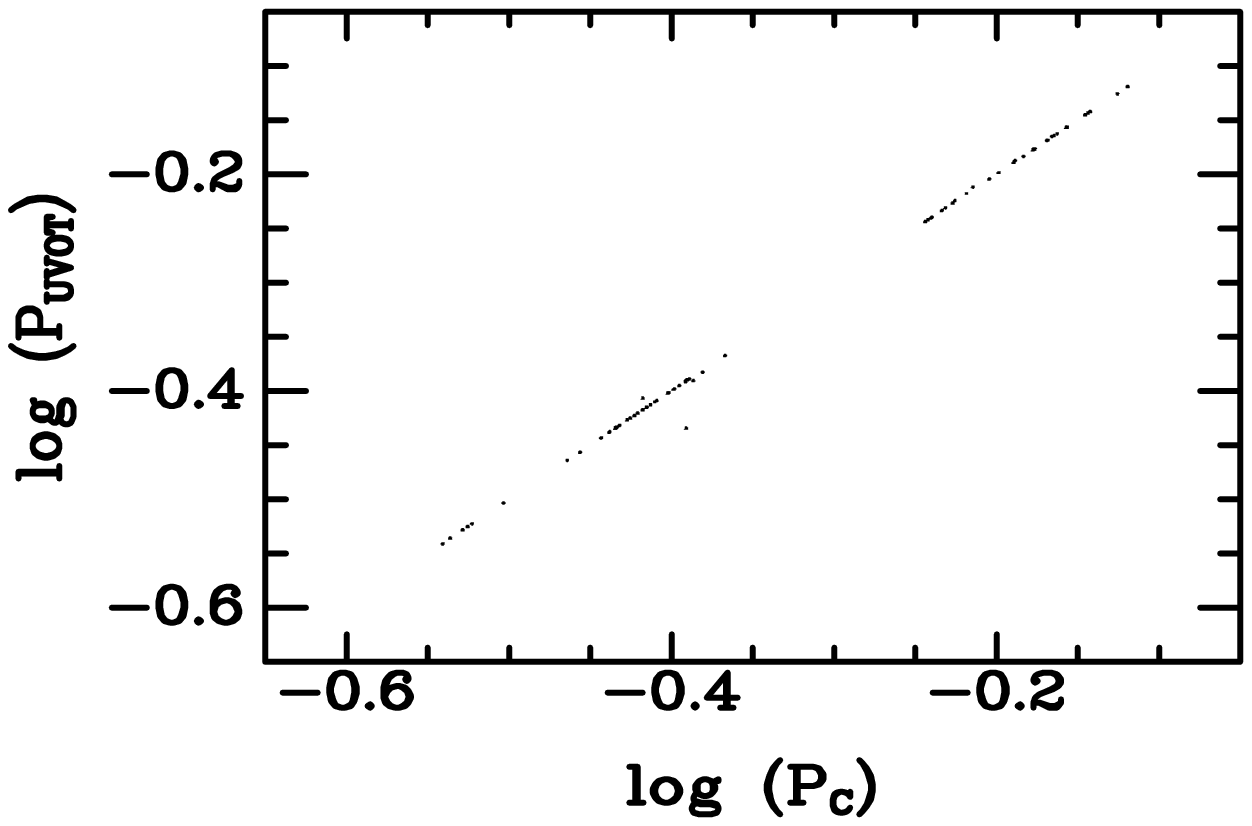}
\end{center}
\caption{A comparison of the periods derived in this study (ordinate) against the compilation of C01 (abscissa) for the globular cluster M~15.  The outliers
are discussed in the text.
\label{f:M15periodcomp}}
\end{figure}

\section{Analysis}
\label{s:analysis}

\subsection{RR Lyrae Star Pulsational Properties}
\label{ss:RRL}

Figure \ref{f:ampcomp} compares the $uvm2$ amplitudes with SHARP values less than 0.5
to the $V$-band amplitudes compiled in C01 (top).  The top panel shows 130 RR Lyrae variable stars in M3 while the bottom shows 58 from M15.
As can be seen, the pulsational amplitudes in the UV are substantially larger, with an average $A_{uvm2}/A_{V}$
ratio of 1.84$\pm$0.41 in M3 and 1.70$\pm$0.46 in M15.  The large NUV amplitudes are expected, given both previous investigations and the astrophysical
properties of RR Lyrae stars, which pulsate over a temperature range where the UV flux can increase dramatically. This ratio appears to be constant
with pulsational period.

\begin{figure}[h]
\begin{center}
\includegraphics[scale=1]{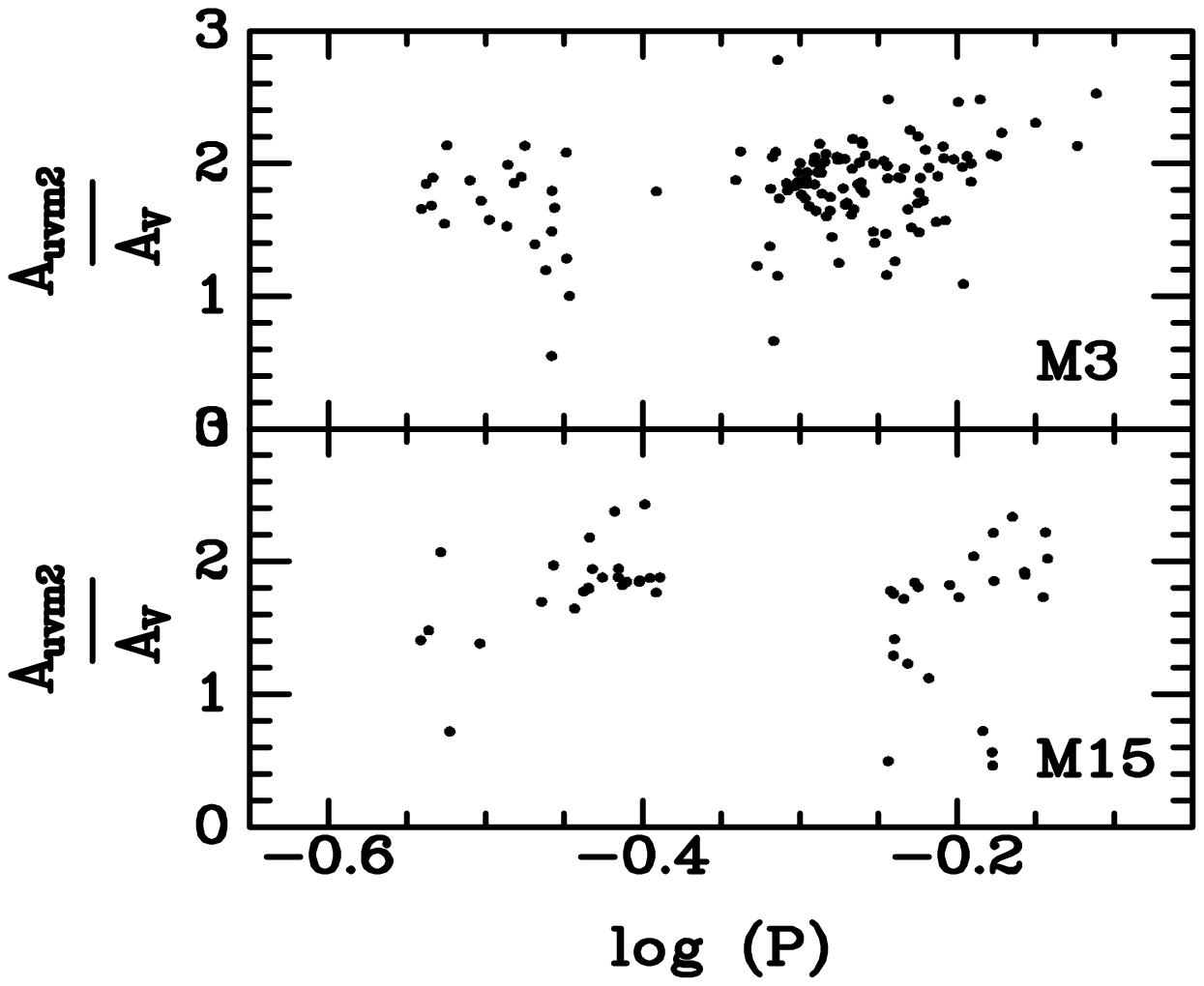}
\end{center}
\caption{A comparison of the $uvm2$ pulsational amplitudes to the $V$ amplitudes collected by C01. Note
that the UV amplitudes are typically much larger than the optical amplitudes, indicating greater sensitivity to temperature changes.\label{f:ampcomp}}
\end{figure}

Figure \ref{f:Bailey} shows the $uvm2$ Bailey (1919) diagram for both clusters.  Large diamonds mark the known double-mode pulsators in M~3 while large squares mark those in M~15.
The familiar loci of RRab stars (the diagonal sequences at long periods) and RRc stars (the relatively
flatter trend at shorter periods) are plainly visible.  We have overlayed fits to the trends of stars in M~3 (solid) and M~15 (dashed) to draw the eye to these sequences.
The comparison between M~3 and M~15 provides
an excellent illustration of the Oosterhoff effect.  Compared to the RR Lyrae stars in M~3, those of M~15 are at longer periods, in both fundamental and first
overtone.  M~15 also shows a much larger proportion of first overtone pulsators.  

\begin{figure}[h]
\begin{center}
\includegraphics[scale=1]{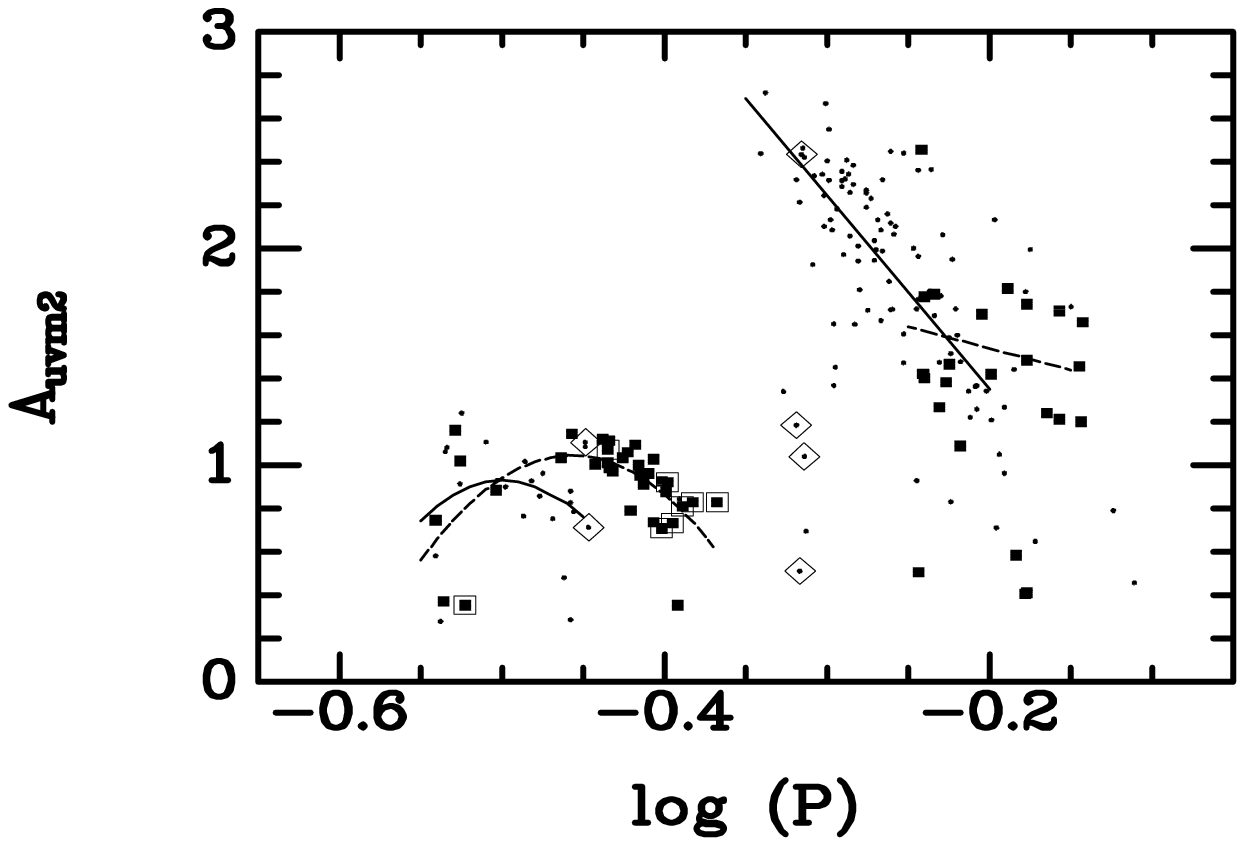}
\end{center}
\caption{A Bailey diagram of period against pulsational amplitudes for the $uvm2$ data for M~3 (points) and M~15 (solid squares).  Open diamonds and squares mark known double-mode pulsators in 
M~3 and M~15, respectively.  The stars
at longer periods ($log(P) > -0.4$) are the fundamental mode RRab pulsators while the stars at shorter periods ($log(P) < -0.4$) are the first overtone RRc pulsators.  The lines
denote fit period-amplitude relations for for M~3 (solid) and M~15 (dashed).\label{f:Bailey}}
\end{figure}

Both clusters' RRab stars follow a linear log $P$-$A_{uvm2}$ sequence, although M~15's RRab
show a much shallower slope.  There have been indications that the RRab locus is slightly curved, but we do not show this clearly in our data.
Interestingly, the RRc stars show a rough diagonal sequence
or possibly even a parabola.  This has been hinted at before in RR Lyrae data (see, for example, the compilation of Bailey diagrams in
Smith et al. 2011) and has been predicted based on theoretical models (see, e.g. Bono et al. 1997).  In this scenario, 
pulsational amplitude linearly tracks temperature for fundamental pulsators but tracks it non-linearly for first overtone pulsators, creating the parabola shape.  M~3's RRc 
log $P$-$A_{uvm2}$ sequence is consistent with a parabola.  However, the RRc stars would also be consistent with a
linear slope or a constant value.  By constrast, M~15's more numerous RRc stars show a distinct parabola shape, with a log $P$-$A_{uvm2}$ sequence that is not consistent with either a
linear slope or a flat trend.  This suggests that NUV surveys of other clusters, especially OoII clusters, could provide a unique test of theoretical models
of first overtone pulsators.

One of the fundamental measures used to study RRab stars is period shift -- the displacement of RRab stars to longer periods at
equal amplitudes in comparison to a reference population (Sandage 1982a, 1982b; Carney et al. 1992).  Star of equal amplitude have similar effective temperatures (Sandage et al. 1981)
so an increase in period at a fixed amplitude is likely due to increased luminosity, according to the pulsation equation of van Albada \& Baker (1971).  The
period shift is postulated to arise from differences in metallicity (see, e.g., Sandage et al. 1981) although other studies have indicated that Oosterhoff
type can play a critical role (see, e.g., Bono et al. 2007).  A fuller discussion of this debate can be found in Smith et al. (2011).

M~3 is the usual reference population used to measure period shift.  We define the fundamental RRab sequence in UVOT magnitudes as:

\begin{center}
$A_{uvm2} = -0.407 - 8.85 \times$ log $P$
\end{center}

The slope of this relation between amplitude and period is steeper in $uvm2$ compared to optical passbands (see, e.g., Siegel \& Majewski 2000, where
the period shift slope for M~3 is calculated as 7.75 and 6.41 in the $B$ and $V$ bands, respectively).  This confirms that the $uvm2$ amplitudes are more sensitive
to pulsational period (and thus, effective temperature) than the optical passbands.

We define the period shift from this relation as:

\begin{center}
$\Delta$ log $P$ = $- (0.0460 + 0.113 A_{uvm2} + $ log $P$ )
\end{center}

Figure \ref{f:shift} shows the period-shift of the RRab stars in both clusters.  The M3 RRab stars cluster around the origin with a median $\Delta$ log $P$ of 0.000,
which is expected since they define the period shift.  The M15 RRab stars show a median period shift of -0.008, although it is weaker for the shorter-period stars (as hinted
at by the shallower slope of the RRab sequence in Figure \ref{f:Bailey}).
This confirms that the period shift seen in Figure \ref{f:Bailey} is real but suggests that further investigation is warranted to confirm that the period-shift
separation between OoI and OoII is as clean in the NUV as it is in the optical passbands.

\begin{figure}[h]
\begin{center}
\includegraphics[scale=1]{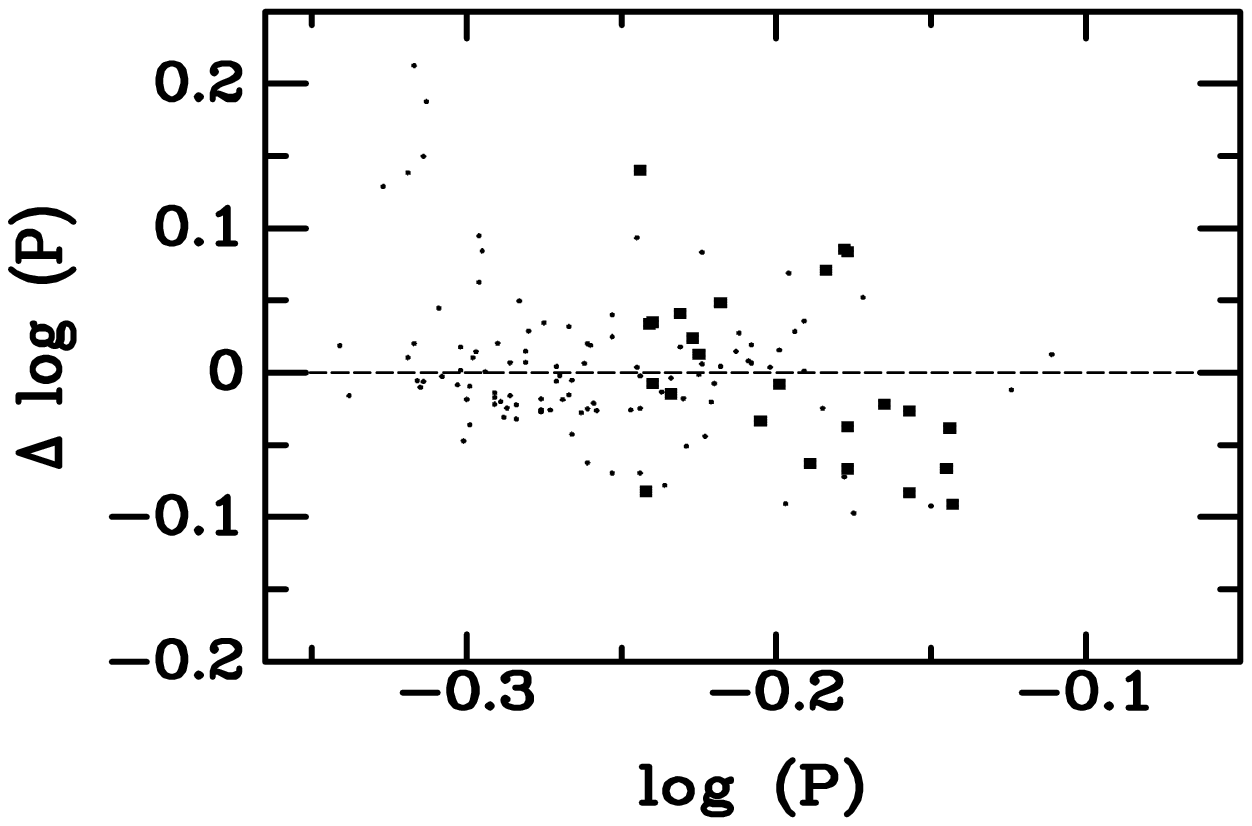}
\end{center}
\caption{The period shift of RRab stars in M3 (points) and M15 (solid squares).  The y-axis represent the displacement of the RRab pulsational period from
the sequence of M~3.\label{f:shift}}
\end{figure}

\subsection{RR Lyrae Star Temperatures and Surface Gravities}

The physical properties of the RR Lyrae stars in M~3 and M~15 have been explored in exhaustive detail by previous investigations.  We do wish
to revisit one aspect of the RR Lyrae stars to highlight something unique in the NUV data: a strong sensitivity to both temperature and surface gravity.

There are a variety of diagnostics that can measure the temperatures of RR Lyrae stars. CCC05 list six different methods that they use
to measure temperatures for 133 RR Lyrae variable stars in M~3, providing a very through investigation of RR Lyrae star temperatures.

Figure \ref{f:temp} compares the temperatures CCC05 calculate from $B-V$ photometry and the temperature scale of Sekiguchi \& Fukugita (2000) to three different photometric
indices: the $B-V$ colors used in that study and $uvm2-B$ and $uvm2-V$ indices derived by combining our photometry with that of CCC05.  We attempted a comparison to UVOT
colors but the $uvw1$ and $uvw2$ passbands had too few observations to derive useful mean magnitudes for the RR Lyrae.

\begin{figure}[h]
\begin{center}
\includegraphics[scale=1]{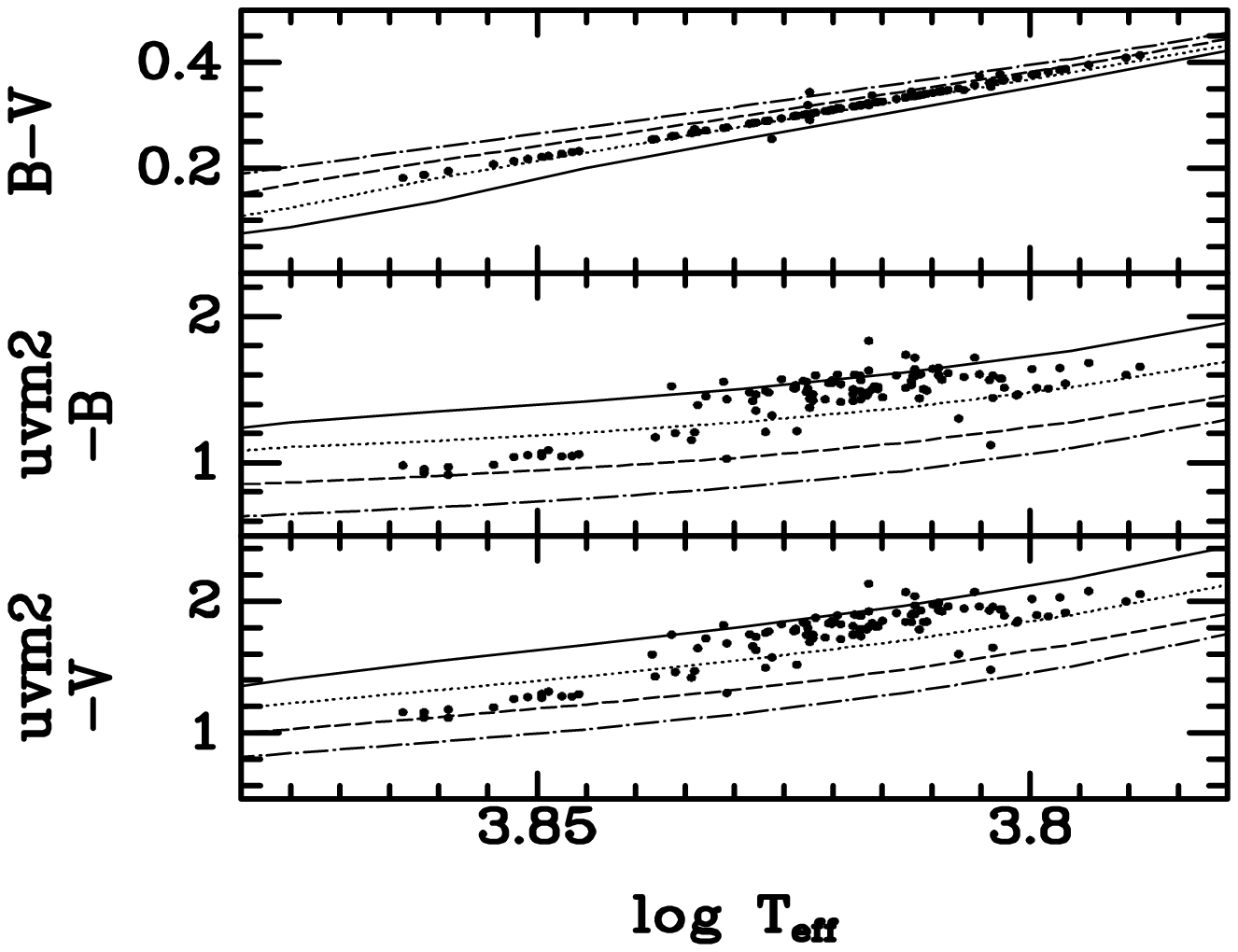}
\end{center}
\caption{A comparison of $B-V$, $uvm2-B$ and $uvm2-V$ photometry to theoretical models of Castelli \& Kurucz (2003).  The lines show temperature-color relations
for $log~g$ values of 2.0 (solid line), 2.5 (dotted), 3.0 (dashed) and 3.5 (dot-dash).\label{f:temp}}
\end{figure}

Overlaid on the panels are predicted theoretical colors as a function of $T_{eff}$ for a range of surface gravities.  The theoretical colors were derived
from the models of Castelli \& Kurucz (2003) using a metallicity of [Fe/H]=-1.5, an $\alpha$-abundance of [$\alpha/Fe$]=+0.4 and microturbulent velocity
of $v_{turb}=2.0$ km s$^{-1}$.
The theoretical magnitudes were corrected for reddening of $E(B-V)=0.01$, using the non-linear corrections for the $uvm2$ photometry discussed
in the Appendix of Paper I.

The top panel of Figure \ref{f:temp} shows the CCC05 colors compared to the theoretical temperatures.  They obviously track well, given that the latter was derived from the former.  Note,
that the broadband photometry is relatively insensitive to the surface gravity.  The lines for $log~g$ values from 2.0 to 3.5 bunch very closely together.

The two lower panels of Figure \ref{f:temp} show the combination of CCC05 photometry and $uvm2$ photometry.  These hybrid UV-optical colors are far more sensitive to the surface gravity.
This is expected, given that the combination of optical and UV filters straddles the Balmer jump, the strength of which is highly sensitive to surface gravity in
moderate to hot stars.  We see that that while the cooler stars have inferred surface gravities in the range of $log~g$ 2.0 to 2.5, the hotter
stars have surface gravities of nearly 3.0.

This is roughly consistent with both observational data and theoretical expectations.  CCC05 calculate $log~g$ from the equation of stellar structure and show a
steady increase with effective temperature from $log~g \sim 2.7$ to $log~g \sim 3.0$.  We generated a synthetic horizontal branch for M~3 using the online tool of
Dotter et al. (2008) and assuming
[Fe/H]=-1.5, [$\alpha$/Fe]=+0.2 and an average mass of 0.8 $M_{\odot}$.  The model shows surface gravity gradually increasing with temperature
over the range of RR Lyrae stars from $log~g=2.67$ at $log (T_{eff})=3.78$ to $log~g=3.11$ at $log (T_{eff})=3.88$.

Our analysis shows the cooler RRab stars having slightly lower surface gravities than expected by about 0.5 dex.  This is likely due to our use of intensity-weighted
mean magnitudes which are, at best, an approximation of the static color an RR Lyrae star would have if it were not pulsating.  Bono et al. (1995) showed that colors
derived from intensity weighted magnitudes are systematically offset from the static-star colors, owing to the effect of increasing amplitude.  They compile a table
of corrections to be applied to $B-V$ colors for fundamental and first overtone colors.  It is not clear how these would scale to the UV and deriving these corrections
is beyond the scope of the current study.  However, the discrepancy between the models and the data in Figure \ref{f:temp} indicates that this correction is likely to be significant (several
tenths of a magnitude) and that further investigation is critical if we are to derive fundamental stellar parameters from NUV-optical photometry.

The dual sensitivity of the NUV photometry to both temperature and surface gravity provides a unique window into the astrophysics of pulsating stars.  
By comparing an index that is not sensitive to surface gravity (such as $B-V$) to one that is ($uvm2-B$ or $uvm2-V$), we can simultaneously measure
both properties from photometry alone.  Our M~3 results demonstrate that this can be done and that is it roughly consistent with theoretical
expectations and previous observational analysis.  This will be a powerful tool for future investigations into other populations of RR Lyrae stars.

\subsection{RR Lyrae Star Period-Luminosity Relationship}

The diagonal HB seen in Figures \ref{f:m3cmd} and \ref{f:m15cmd} would suggest a potential period-luminosity relationship for RR Lyrae stars in the NUV.  The
top panel of figure \ref{f:pl} shows
the period-magnitude relation for both clusters and appears to confirm this supposition.  The stars follow a sequence of increasing magnitude
with period for both populations.

\begin{figure}[h]
\begin{center}
\includegraphics[scale=1]{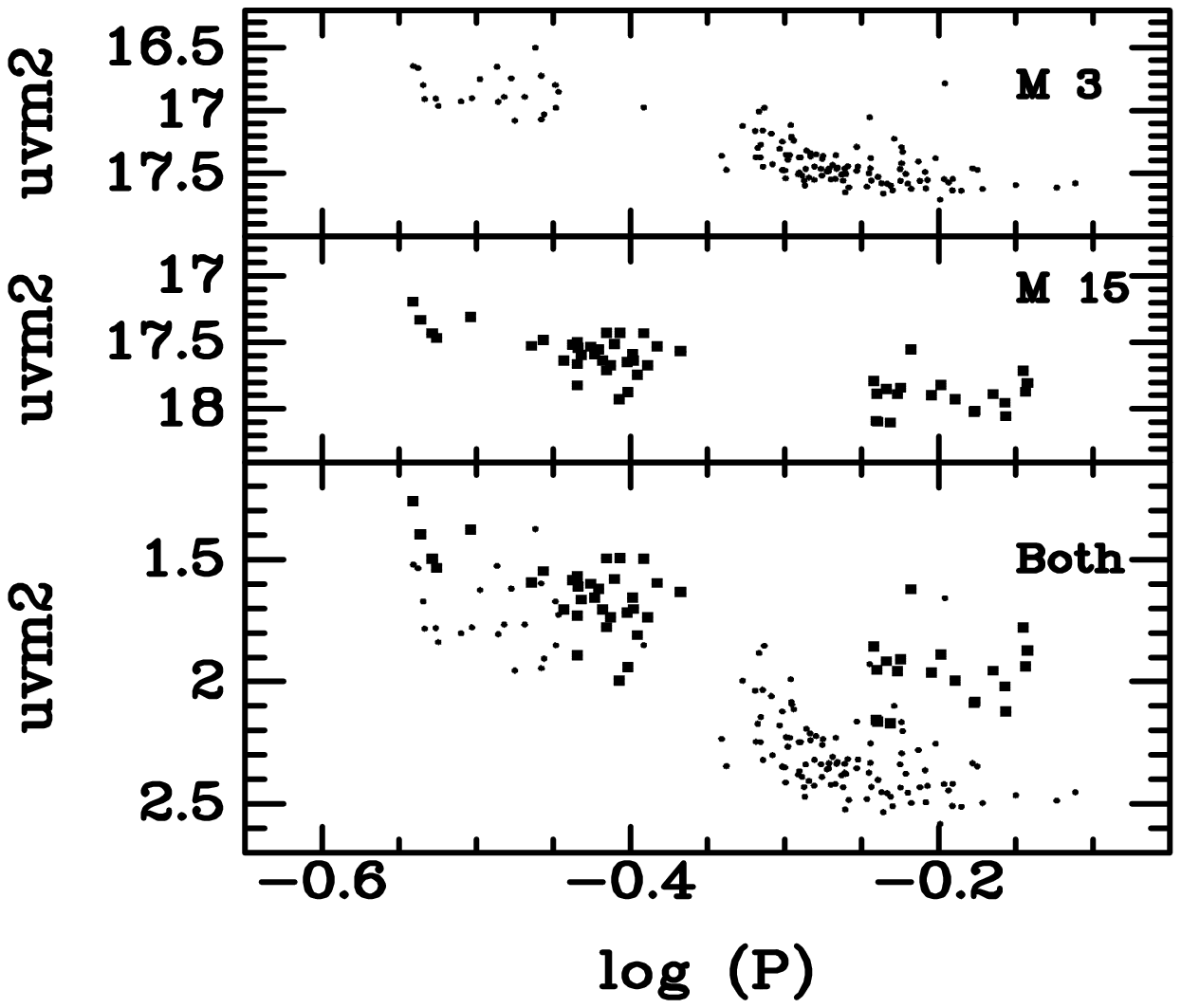}
\end{center}
\caption{The period-luminosity relation of RR Lyrae stars in M3 (top) and M15 (middle).  The outliers are likely blends.  The bottom panel shows both clusters
corrected for distance modulus and reddening.\label{f:pl}}
\end{figure}

Two of M~3's stars are substantially brighter than the period-luminosity relationship by more than half a magnitude.  This might suggest they are
foreground or evolved objects but their
$V$ and $B$ magnitudes in C01 are well within the primary
sequence.  Although their SHARP values are below 0.5, they are slightly high ($\sim0.3$) and their amplitudes are a bit low compared to stars of similar period.  This
suggests they may be blends.  Removing these two stars for consideration, we find a linear fit to the period-magnitude sequence of:

\begin{center}
$uvm2 = 18.06 + 2.41 \times$ log $P$
\end{center}

Correcting for the distance modulus and reddening listed in Harris (1996) yields:

\begin{center}
$M_{uvm2} = 2.93 + 2.41 \times$ log $P$ 
\end{center}

M~15 has five stars that are outliers by more than half a magnitude on the period-luminosity relationship, all of which have elevated SHARP values (0.22-0.43).  Removing
these from the fit yields:

\begin{center}
$uvm2 = 18.18 + 1.42 \times$ log $P$
\end{center}

Correcting for the distance modulus and reddening listed in Harris (1996) yields:

\begin{center}
$M_{uvm2} = 2.25 + 1.42 \times$ log $P$
\end{center}

The bottom panel of Figure \ref{f:pl} shows the stars from both M~3 and M~15, corrected for the distances modulus and foreground reddening values listed in Harris (1996).  As can be seen, the RR Lyrae
in M~15 are brighter than those of comparable period in M3.  There is some scatter and non-linearity in the comparison, with the coolest RRab stars being
almost half a magnitude brighter while the hottest RRc stars are a few tenths of a magnitude brighter (as implied by the different slopes in the period-magnitude
relations of the two clusters).  The difference is dependent on the assumed reddening (for most of the RR Lyrae temperature range, the extinction in $uvm2$ is greater than
8.0).  However, even modulo reddening, this comparison suggests that the NUV period-luminosity relationship has a strong metallicity dependence, far stronger than
is seen in the IR, where the metallicity dependence is a few hundredths of a magnitude to a few tenths of a magnitude (Bono et al. 2001; Sollima et al. 2006; 
Muraveva et al. 2015).  Alternatively, the different slope in the period-magnitude relationships could indicate that the relationship changes depending on Oosterhoff
type.

As simple fit with a cross-term to these two trends yields the relation:

\begin{center}
$M_{uvm2} = 4.103 + 0.782$ [Fe/H] + $4.117$ log $P + 1.138$ log $P$ [Fe/H]
\end{center}

\noindent confirming both the strong metallicity dependence and the non-linearity.  However, this "fit" is based on two clusters.  Many more clusters
are needed to determine if the period-magnitude-metallicity relationship varies smoothly with metallicity or is different for the different Oosterhooff classifications.
It would also be useful to study objects that have a range of metallicities, such as the Magellanic Clouds or Omega Centauri, so that distance and reddening differences
can be minimized.  The most robust evaluation would come from studying RR Lyrae with trigonometric parallaxes, such as those expected to emerge from the GAIA mission.

%\subsection{Other variable stars}

\section{Conclusions}
\label{s:conc}

We have presented the first thorough investigation into the NUV pulsational properties of variable stars in the well-studied globular
clusters M~3 and M~15.  We cross-identify and fit periods to 201 variable stars in M~3 and 85 in M~15.  Of these 286 stars, 282 are RR Lyrae stars.  The remainder
are three Cepheids and one SX Phoenicis star.  We identify one potential new variable in M~3 and one new RRc star in M~15.

We find that the RR Lyrae stars have, as expected, very large NUV amplitudes compared to the optical and infrared, with value ranging up to 2.7 magnitudes.  We find that the RR Lyrae
conform to the same paradigms established in the optical, exhibiting a clear Oosterhoff effect, a period shift between M~3 and M~15 and a correlation between
temperature and pulsational amplitude.  We show that the combination of NUV and optical data shows remarkable sensitivity to both the temperature and surface
gravity of the RR Lyrae stars, although our comparison to the Kurucz models underestimates the surface gravity of the cooler RRab stars likely due to the effect
of the large amplitudes on the intensity-weighted $uvm2-B$ and $uvm2-V$ colors.

We further find that the stars exhibit a clear period-magnitude relation in the NUV.  While we only have two cluster to compare, the difference between them
suggests that the period-lumonisity relationship has a strong and possibly non-linear dependence on metallicity.

Swift continues to observe Galactic clusters that are known to host populations of RR Lyrae, both as part of our ongoing investigation and as part of routine
monitoring of X-ray transients in globular clusters (Bahramian et al. 2013).  This will enable more through explorations of the interaction of RR Lyrae pulsational
properties, metallicity and reddening.

\acknowledgements

The authors acknowledge sponsorship at PSU by NASA contract NAS5-00136.  This research was also supported by the Swift Guest Investigator
Program through NASA grant NNX14AC31G.  The authors thank Caryl Gronwall for useful comments and the anonymous referee for useful
corrections.  The Institute for Gravitation and the Cosmos is supported by the Eberly College of Science and the Office of the Senior Vice President
for Research at the Pennsylvania State University.

\bibliographystyle{apj}

\begin{thebibliography}{}
\bibitem[Bahramian et al.(2013)]{2013ATel.5396....1B} Bahramian, A., 
Heinke, C.~O., Sivakoff, G.~R., Altamirano, D., 
\& Wijnands, R.\ 2013, The Astronomer's Telegram, 5396, 1 
\bibitem[Bailey et al.(1919)]{1919AnHar..78..195B} Bailey, S.~I., Leland, 
E.~F., Woods, I.~E., 
\& Pickering, E.~C.\ 1919, Annals of Harvard College Observatory, 78, 195 
\bibitem[Baker 
\& Baker(1956)]{1956AJ.....61..283B} Baker, R.~H., \& Baker, H.~V.\ 1956, \aj, 61, 283 
\bibitem[Bakos et al.(2000)]{2000AcA....50..221B} Bakos, G.~A., Benko, 
J.~M., \& Jurcsik, J.\ 2000, \actaa, 50, 221 
\bibitem[Benk{\H o} et al.(2006)]{2006MNRAS.372.1657B} Benk{\H o}, J.~M., 
Bakos, G.~{\'A}., \& Nuspl, J.\ 2006, \mnras, 372, 1657 
\bibitem[Bingham et al.(1984)]{1984MNRAS.209..765B} Bingham, E.~A., 
Cacciari, C., Dickens, R.~J., \& Pecci, F.~F.\ 1984, \mnras, 209, 765 
\bibitem[Blazhko (1907)]{1907AN...175..325} Blazhko, S. 1907, {\it Astron. Nachr.}, 175, 325
\bibitem[Bonnell et al.(1982)]{1982PASP...94..910B} Bonnell, J., Wu, C.-C., 
Bell, R.~A., \& Hutchinson, J.~L.\ 1982, \pasp, 94, 910 
\bibitem[Bono et al.(1995)]{1995ApJS...99..263B} Bono, G., Caputo, F., 
\& Stellingwerf, R.~F.\ 1995, \apjs, 99, 263
\bibitem[Bono et 
al.(1997)]{1997A&AS..121..327B} Bono, G., Caputo, F., Castellani, V., \& Marconi, M.\ 1997, \aaps, 121, 327 
\bibitem[Bono et al.(2001)]{2001MNRAS.326.1183B} Bono, G., Caputo, F., 
Castellani, V., Marconi, M., \& Storm, J.\ 2001, \mnras, 326, 1183
\bibitem[Bono et al. (2007)]{2007AA.476.779} Bono, G., Caputo, F., \& Di Criscienzo, M. 2007, \aap, 476, 779
\bibitem[Breeveld et al.(2010)]{2010MNRAS.tmp..874B} Breeveld, A.~A., et 
al.\ 2010, \mnras, 874 
\bibitem[Breeveld et al.(2011)]{2011AIPC.1358..373B} Breeveld, A.~A., 
Landsman, W., Holland, S.~T., et al.\ 2011, American Institute of Physics 
Conference Series, 1358, 373 
\bibitem[Burrows et al.(2005)]{2005SSRv..120..165B} Burrows, D.~N., et al.\ 
2005, Space Science Reviews, 120, 165
\bibitem[Butler et al.(1998)]{1998MNRAS.296..379B} Butler, R.~F., Shearer, 
A., Redfern, R.~M., et al.\ 1998, \mnras, 296, 379 
\bibitem[Cacciari et al.(2005)]{2005AJ....129..267C} Cacciari, C., Corwin, 
T.~M., \& Carney, B.~W.\ 2005, \aj, 129, 267 [CCC05]
\bibitem[Carney et al.(1992)]{1992ApJ...386..663C} Carney, B.~W., Storm, 
J., \& Jones, R.~V.\ 1992, \apj, 386, 663 
\bibitem[Carretta et al.(1998)]{1998MNRAS.298.1005C} Carretta, E., 
Cacciari, C., Ferraro, F.~R., Fusi Pecci, F., 
\& Tessicini, G.\ 1998, \mnras, 298, 1005 
\bibitem[Castelli 
\& Kurucz(2003)]{2003IAUS..210P.A20C} Castelli, F., \& Kurucz, R.~L.\ 2003, Modelling of Stellar Atmospheres, 210, 20P
\bibitem[Clement et al.(2001)]{2001AJ....122.2587C} Clement, C.~M., Muzzin, 
A., Dufton, Q., et al.\ 2001, \aj, 122, 2587 [C01]
\bibitem[Corwin 
\& Carney(2001)]{2001AJ....122.3183C} Corwin, T.~M., \& Carney, B.~W.\ 2001, \aj, 122, 3183 [CC)1]
\bibitem[Corwin et al.(2008)]{2008AJ....135.1459C} Corwin, T.~M., 
Borissova, J., Stetson, P.~B., et al.\ 2008, \aj, 135, 1459 [C08]
\bibitem[Del Principe et al.(2005)]{2005AJ....129.2714D} Del Principe, M., 
Piersimoni, A.~M., Bono, G., et al.\ 2005, \aj, 129, 2714 
\bibitem[Dieball et al.(2007)]{2007ApJ...670..379D} Dieball, A., Knigge, 
C., Zurek, D.~R., et al.\ 2007, \apj, 670, 379 
\bibitem[Dotter et al.(2008)]{2008ApJS..178...89D} Dotter, A., Chaboyer, 
B., Jevremovi{\'c}, D., et al.\ 2008, \apjs, 178, 89 
\bibitem[Downes et al.(2004)]{2004AJ....128.2288D} Downes, R.~A., Margon, 
B., Homer, L., \& Anderson, S.~F.\ 2004, \aj, 128, 2288 
\bibitem[Ferraro 
\& Paresce(1993)]{1993AJ....106..154F} Ferraro, F.~R., \& Paresce, F.\ 1993, \aj, 106, 154 
\bibitem[Filippenko 
\& Simon(1981)]{1981AJ.....86..671F} Filippenko, A.~V., \& Simon, R.~S.\ 1981, \aj, 86, 671 
\bibitem[Gehrels et al.(2004)]{2004ApJ...611.1005G} Gehrels, N., et al.\ 2004, \apj, 611, 1005
\bibitem[Guhathakurta et al.(1994)]{1994AJ....108.1786G} Guhathakurta, P., 
Yanny, B., Bahcall, J.~N., \& Schneider, D.~P.\ 1994, \aj, 108, 1786 
\bibitem[Hardie(1955)]{1955ApJ...122..256H} Hardie, R.~H.\ 1955, \apj, 122, 
256 
\bibitem[Harris(1996)]{1996AJ....112.1487H} Harris, W.~E.\ 1996, \aj, 112, 
1487 
\bibitem[Hartman et al.(2005)]{2005AJ....129.1596H} Hartman, J.~D., 
Kaluzny, J., Szentgyorgyi, A., \& Stanek, K.~Z.\ 2005, \aj, 129, 1596 
\bibitem[Hutchinson et al.(1977)]{1977ApJ...211..207H} Hutchinson, J.~L., 
Lillie, C.~F., \& Hill, S.~J.\ 1977, \apj, 211, 207 
\bibitem[Kaluzny et al.(1998)]{1998MNRAS.296..347K} Kaluzny, J., Hilditch, 
R.~W., Clement, C., \& Rucinski, S.~M.\ 1998, \mnras, 296, 347 
\bibitem[Kov{\'a}cs 
\& Kupi(2007)]{2007A&A...462.1007K} Kov{\'a}cs, G., \& Kupi, G.\ 2007, \aap, 462, 1007 
\bibitem[Layden(1998)]{1998AJ....115..193L} Layden, A.~C.\ 1998, \aj, 115, 
193 
\bibitem[Muraveva et al.(2015)]{2015ApJ...807..127M} Muraveva, T., Palmer, 
M., Clementini, G., et al.\ 2015, \apj, 807, 127 
\bibitem[Neeley et al.(2015)]{2015ApJ...808...11N} Neeley, J.~R., Marengo, 
M., Bono, G., et al.\ 2015, \apj, 808, 11
\bibitem[Oosterhoff(1939)]{1939Obs....62..104O} Oosterhoff, P.~T.\ 1939, 
The Observatory, 62, 104 
\bibitem[Pei(1992)]{Pei92} Pei, Y. C., 1992, \apj, 395, 130
\bibitem[Poole et al.(2008)]{2008MNRAS.383..627P} Poole, T.~S., et al.\ 
2008, \mnras, 383, 627 
\bibitem[Popielski et al.(2000)]{2000AcA....50..491P} Popielski, B.~L., 
Dziembowski, W.~A., \& Cassisi, S.\ 2000, \actaa, 50, 491 
\bibitem[Roming et al.(2000)]{2000SPIE.4140...76R} Roming, P.~W., Townsley, 
L.~K., Nousek, J.~A., et al.\ 2000, \procspie, 4140, 76 
\bibitem[Roming et al.(2004)]{2004SPIE.5165..262R} Roming, P.~W.~A., 
Hunsberger, S.~D., Mason, K.~O., et al.\ 2004, \procspie, 5165, 262 
\bibitem[Roming et al.(2005)]{2005SSRv..120...95R} Roming, P.~W.~A., et 
al.\ 2005, Space Science Reviews, 120, 95  
\bibitem[Rosino(1950)]{1950ApJ...112..221R} Rosino, L.\ 1950, \apj, 112, 
221 
\bibitem[Rosino(1969)]{1969IBVS..327....1R} Rosino, L.\ 1969, Information 
Bulletin on Variable Stars, 327, 1 
\bibitem[Sandage et al.(1981)]{1981ApJS...46...41S} Sandage, A., Katem, B., 
\& Sandage, M.\ 1981, \apjs, 46, 41 
\bibitem[Sandage(1982a)]{1982ApJ...252..553S} Sandage, A.\ 1982, \apj, 252, 
553 
\bibitem[Sandage(1982b)]{1982ApJ...252..574S} Sandage, A.\ 1982, \apj, 252, 
574
\bibitem[Sandquist et al.(2010)]{2010AJ....139.2374S} Sandquist, E.~L., 
Gordon, M., Levine, D., \& Bolte, M.\ 2010, \aj, 139, 2374 
\bibitem[Schiavon et al.(2012)]{2012AJ....143..121S} Schiavon, R.~P., 
Dalessandro, E., Sohn, S.~T., et al.\ 2012, \aj, 143, 121 
\bibitem[Sekiguchi 
\& Fukugita(2000)]{2000AJ....120.1072S} Sekiguchi, M., \& Fukugita, M.\ 2000, \aj, 120, 1072 
\bibitem[Siegel et al.(2014)]{2014AJ....148..131S} Siegel, M.~H., 
Porterfield, B.~L., Linevsky, J.~S., et al.\ 2014, \aj, 148, 131 [Paper I]
\bibitem[Siegel 
\& Majewski(2000)]{2000AJ....120..284S} Siegel, M.~H., \& Majewski, S.~R.\ 2000, \aj, 120, 284 
\bibitem[Silbermann 
\& Smith(1995)]{1995AJ....110..704S} Silbermann, N.~A., \& Smith, H.~A.\ 1995, \aj, 110, 704
\bibitem[Smith(1995)]{1995Sci...270.1236S} Smith, H.~A.\ 1995, Science, 
270, 1236 
\bibitem[Smith et al.(2011)]{2011rrls.conf...17S} Smith, H.~A., Catelan, 
M., 
\& Kuehn, C.\ 2011, RR Lyrae Stars, Metal-Poor Stars, and the Galaxy, 17 
\bibitem[Sollima et al.(2006)]{2006MNRAS.372.1675S} Sollima, A., Cacciari, 
C., \& Valenti, E.\ 2006, \mnras, 372, 1675 
\bibitem[Stetson(1987)]{1987PASP...99..191S} Stetson, P.~B.\ 1987, \pasp, 
99, 191 
\bibitem[Stetson(1994)]{1994PASP..106..250S} Stetson, P.~B.\ 1994, \pasp, 
106, 250
\bibitem[Thomson et al.(2010)]{2010MNRAS.406.1084T} Thomson, G.~S., 
Dieball, A., Knigge, C., Long, K.~S., 
\& Zurek, D.~R.\ 2010, \mnras, 406, 1084 
\bibitem[Tuairisg et al.(2003)]{2003MNRAS.345..960T} Tuairisg, S.~{\'O}., 
Butler, R.~F., Shearer, A., et al.\ 2003, \mnras, 345, 960 
\bibitem[van Albada \& Baker (1971)]{1971ApJ.169.311} van Albada, T.S. \& Baker, N. 1971, \apj, 169, 311
\bibitem[Wheatley et al.(2005)]{2005ApJ...619L.123W} Wheatley, J.~M., 
Welsh, B.~Y., Siegmund, O.~H.~W., et al.\ 2005, \apjl, 619, L123 
\bibitem[Wheatley et al.(2012)]{2012PASP..124..552W} Wheatley, J., Welsh, 
B.~Y., \& Browne, S.~E.\ 2012, \pasp, 124, 552 
\bibitem[Yang 
\& Sarajedini(2012)]{2012MNRAS.419.1362Y} Yang, S.-C., \& Sarajedini, A.\ 2012, \mnras, 419, 1362 
\bibitem[Zheleznyak \& Kravtsov(2003)]{2003AstL...29..599Z} Zheleznyak, A.~P., \& Kravtsov, V.~V.\ 2003, Astronomy Letters, 29, 599 
Query Results from the ADS Database


\end{thebibliography}

\end{document}